\documentclass[fleqn,usenatbib]{mnras}

\usepackage{newtxtext,newtxmath}
\usepackage{longtable}
\usepackage{multirow}

\usepackage[T1]{fontenc}
\DeclareRobustCommand{\VAN}[3]{#2}
\let\VANthebibliography\thebibliography
\def\thebibliography{\DeclareRobustCommand{\VAN}[3]{##3}\VANthebibliography}

\usepackage{graphicx}	% Including figure files
\usepackage{amsmath}	% Advanced maths commands
\usepackage{siunitx}
\usepackage{hyperref}
\DeclareSIUnit\erg{erg}
\title[Interstellar Comet 2I/Borisov]{Pre-perihelion Monitoring of Interstellar Comet 2I/Borisov}

\author[G. P. Prodan et al.]{
George P. Prodan,$^{1}$\thanks{E-mail: prodangp9@gmail.com}
Marcel Popescu,$^{1,2}$
Javier Licandro,$^{3,4}$
Mohammad Akhlaghi,$^{5}$
Julia de León,$^{3,4}$
\newauthor
Eri Tatsumi,$^{6}$
Bogdan Adrian Pastrav,$^{7}$
Jacob M. Hibbert,$^{8}$
Ovidiu Văduvescu,$^{8,3,2}$
\newauthor
Nicolae Gabriel Simion,$^{1}$
Enric Pallé,$^{3,4}$
Norio Narita,$^{9,10,3}$
Akihiko Fukui, $^{9,3}$
Felipe Murgas$^{3,4}$
\\ \\
$^{1}${Astronomical Institute of the Romanian Academy, 5 Cuțitul de Argint, 040557 Bucharest, Romania}\\
$^{2}${Faculty of Sciences, University of Craiova, Craiova, Romania
}\\
$^{3}${Instituto de Astrofísica de Canarias (IAC), C/Vía Láctea s/n, E-38205, La Laguna, Spain
}\\
$^{4}${Departamento de Astrofísica (ULL), E-38205, La Laguna, Spain
}\\
$^{5}${Centro de Estudios de F\'isica del Cosmos de
Arag\'on (CEFCA), Plaza San Juan 1, 44001 Teruel, Spain}\\
$^{6}${Institute of Space and Astronautical Science, Japan Aerospace Exploration Agency (JAXA), Kanagawa, Japan
}\\
$^{7}${Institute of Space Science, Atomiștilor 409, 077125, Bucharest-Măgurele, Romania
}\\
$^{8}${Isaac Newton Group of Telescopes (ING), Apto. 321, E-38700 Santa Cruz de la Palma, Canary Islands, Spain
}\\
$^{9}$ {Komaba Institute for Science, The University of Tokyo, 3-8-1 Komaba, Meguro, Tokyo 153-8902, Japan}\\
$^{10}${Astrobiology Center, 2-21-1 Osawa, Mitaka, Tokyo 181-8588, Japan}\\
}
\date{Accepted 2024 February 12. Received 2024 February 10; in original form 2024 January 15}

\pubyear{2024}

\begin{document}
\label{firstpage}
\pagerange{\pageref{firstpage}--\pageref{lastpage}}
\maketitle

\begin{abstract}
The discovery of interstellar comet 2I/Borisov offered the unique opportunity to obtain a detailed analysis of an object coming from another planetary system, and leaving behind material in our interplanetary space. We continuously observed 2I/Borisov between October 3 and December 13, 2019 using the 1.52-m Telescopio Carlos S\'{a}nchez equipped with MuSCAT2 instrument, and the 2.54-m Isaac Newton Telescope with Wide Field Camera. We characterize its morphology and spectro-photometric features using the data gathered during this extended campaign. Simultaneous imaging in four bands ($g$, $r$, $i$, and $z_s$) reveals a homogeneous composition and a reddish hue, resembling Solar System comets, and as well a diffuse profile exhibiting familiar cometary traits. We discern a stationary trend fluctuating around a constant activity level throughout October and November 2019. Subsequently, a reduction in activity is observed in December. Dust production and mass loss calculations indicate approximately an average of 4 kg/s before perihelion, while after perihelion the net mass loss is about 0.6 kg/s. Our simulations indicate the most probable size of coma dust particles should be in the range 200-250 nm, and the terminal speed around 300 m/s. The spectrum acquired with the 4.2-m William Herschel Telescope shows the presence of a strong CN line for which we find a gas production rate of $1.2 \times 10^{24}~s^{-1}$. We also detected NH$_2$ and OI bands. The ratio between NH$_2$ and CN productions is $\log (NH_2/CN) =-0.2$. Overall, this observing campaign provides a new understanding of 2I/Borisov's unique characteristics and activity patterns.

\end{abstract}


%%%%%%%%%%%%%%%%%%%%%%%%%%%%%%%%%%%%%%%%%%%%%%%%%%

%%%%%%%%%%%%%%%%% BODY OF PAPER %%%%%%%%%%%%%%%%%%

\section{Introduction}
Comets are ancient celestial objects that offer insight into the origin of the Solar System. The recent discovery of 2I/Borisov, the first known interstellar comet \citep{2021SoSyR..55..124B} and the second interstellar object after 1I/'Oumuamua \citep{Trilling_2018}, has captured the attention of astronomers as a witness of the formation of other planetary systems. The orbit of this comet has led to various proposals regarding its origin, such as being a stardust comet due to its water abundance \citep{2019arXiv191212730E}, or originating from the Wolf 630 dynamical stream or the Ross 573 stellar system \citep{2020A&A...634A..14B}. Also, it was found by simulations that its velocity is very similar to those of the stars located in the vicinity of the Sun \citep{2020MNRAS.495.2053D}. 

Comet 2I/Borisov was discovered by G. Borisov at the Crimean Astrophysical Observatory on August 30, 2019 \citep{mpclink}. At that time, the comet was located at the geocentric distance $\Delta=2.98$ au following a hyperbolic orbit with an eccentricity of 3.35, and an apparent  magnitude in the $V$-band of roughly $V\approx18$. According to the pre-discovery observations \citep{Ye_2020}, the comet was already active at 5 -- 7 au with the onset around 8 au in December-November 2018, suggesting the presence of species that are more volatile than the water, such as CO or CO$_2$. 

Many groups investigated the spectro-photometric properties of 2I/Borisov. First results from 2019 already highlighted some similarities of Borisov with the Solar System comets \citep{Jewitt_2019, Yang_2020, Ye_2020, sekanina20192iborisov, Guzik_2019,2020MNRAS.495.2053D, Bolin_2020}. Not long after, high resolution images allowed for a better estimation of the nucleus radius, between 0.2 and 0.5 km, as reported by \cite{2020ApJ...888L..23J}. The data gathered with Hubble Space Telescope (HST) on multiple observation runs made between November 2019 and January 2020 revealed a dust tail oriented between the anti-solar direction and the negative heliocentric velocity vector, with a subtle inner coma asymmetry \citep{Kim_2020}. On the other hand, the HST images taken in September and October 2019 \citep{Jewitt_2019,Manzini_2020}, showed that the dust was scattered mostly towards the anti-solar direction. \cite{Manzini_2020} argued that the released dust becomes dispersed at a relatively short distance from the nucleus and quickly undergoes deflection due to solar radiation pressure, eventually merging into the tail. This suggests a low ejection speed, possibly associated with the prevalence of smaller-sized dust particles. Also, they suggested that the morphology of 2I is similar to that of comet C/2014 B1 exhibiting similar activation processes as the native comets of the Solar System.

The high-resolution interferometric observations of 2I/Borisov made with the Atacama Large Millimeter/submillimeter Array (ALMA) indicate the presence of compact pebbles of 1 mm radii \cite{Yang_2021}. Also, \cite{10.1093/mnras/staa4022} found that the Mg-Fe conglomerates and organic dust components are predominant in the 2I coma. \cite{opitom_2021_b} detected similarities to Solar System comets in several bands, including FeI and NiI, presenting an abundance ratio of $\log$ (Ni/Fe) = 0.21 $\pm$ 0.18. The dust production rate and mass loss have been computed in several studies. However, different results have been obtained because of the use of various models or assumptions. Implementation of Monte Carlo dust tail models \citep{2020MNRAS.495.2053D, Kim_2020, Manzini_2020} showed that the size distribution of the tail dust particles is characterized by a mean value around 100 $\mu$m - 1 mm, whereas smaller particles are more likely to be found closer to the nucleus.

Regarding the spectroscopy, even if 2I/Borisov has properties similar to those of comets in our Solar System, it exhibits an unusual high ratio of CO with respect to water and HCN \citep{2020NatAs...4..867B, Cordiner_2020}. This aspect is interesting as 2I is considered the most pristine comet ever observed, with polarimetric properties that suggest that it has not interacted with any object since its formation \citep{nature_polarimetry}, maintaining its original characteristics. However, active comets in the Solar System are usually rich in pristine material.

Indeed, no absorption bands of water ice were found at 1.5 and 2 $\mu$m \citep{Yang_2020} in September -- October 2019. On the other hand, strong CN lines at 388 nm were detected by \cite{Fitzsimmons_2019, 2020MNRAS.495.2053D, Aravind_2021}, HCN being suspected as the parent molecule. 

Another spectral feature is the C$_2$ depletion before reaching perihelion, as proved by several groups \citep{Lin_2020,opitom_2019,bannister2020interstellar, kareta_2020}. Other species were also detected, like NH$_2$, for which \cite{bannister2020interstellar} found large abundances compared to those found for Solar System comets. Furthermore, high spectral resolution measurements by \cite{McKay_2020} proved the presence of the forbidden oxygen lines, specifically the O($^1$D) line at 630 nm. This line can be an excellent tracker for water, as it releases OI into the coma by the photo-dissociation of the parent molecule, which is usually H$_2$O \citep{2004come.book..425F}. They found the gas production rate of water $Q(H_2O)=(6.3 \pm 1.5) \times 10^{26} s^{-1}$ on November 10, 2019. The abundance of the parent species depends on the flux ratio between the oxygen's green line at 557 nm and red doublet lines at 630 nm and 636 nm, denoted usually by $G/R$. The other potential parent species, such as CO or CO$_2$, can be neglected if they are not dominant species in the composition. However, this is not the case for 2I, as CO was found to be dominant. \cite{opitom_2021_b} found $G/R$ values ranging from 0.3 in November 2019 - February 2020 to 0.6 in February -- March 2020, showing that CO is more probable to be responsible for the forbidden oxygen lines found in the spectrum of 2I. 

Another remark on the 2I spectrum of the coma is its similarity to the spectra of D-type asteroids, as it was first suggested by \cite{deLeon2019}, and later confirmed by other authors \citep{Yang_2020, 2020MNRAS.495.2053D}. The above mentioned features of the comet, regarding the composition, seem to agree with this idea. The D-type asteroids are characterized by featureless spectra and are supposed to contain organics and volatiles. Tagish Lakes is one of the meteorite spectral analogues for these bodies \citep[e.g.][]{Barucci_2018, 2021Icar..36314295G}.They are considered the most primitve among the asteroids population.

This work focuses on the analysis of spectro-photometric data collected with the 1.52-m Carlos S\'{a}nchez Telescope (TCS), the 2.53-m Isaac Newton Telescope (INT), and the spectral data obtained with the 4.2-m William Herschel Telescope (WHT). The observations were performed during a two and a half months observational campaign between October 3 - December 13, 2019. The data covered the perihelion passage of 2I/Borisov on December 8, 2019 at a distance of $q$ = 2.0065 au (according to NASA JPL Small-Body Database Browser ephemerides\footnote{\url{https://ssd.jpl.nasa.gov/horizons/}}). The goals of this study are to determine the comet's morphology and spectrophotometry, as well as to calculate the dust production rate and the net mass loss of cometary dust. This article is organized as follows. In Section~2 we present the observational campaign of 2I/Borisov. Section~3 describes the data reduction and the analysis method (the mathematical models used are shown in Appendix~A and Appendix~B). The results are presented in Section~4. The discussion and conclusions are provided in Section~5.

\section{Observations}

Our observing campaign aimed to monitor the activity of 2I/Borisov using broad-band filters. The most convenient instrument for this was the MuSCAT2 (Multicolor Simultaneous Camera for studying Atmospheres of Transiting exoplanets), mounted on the 1.52-m TCS, which allows simultaneous imaging with four broad band filters \citep{2019JATIS...5a5001N}. We complemented our observations with spectro-photometric images obtained with the 2.54-m INT telescope. Also, we succeeded to acquire a spectrum with the WHT telescope. These instruments are briefly described bellow.

\subsection{The 1.52-m Carlos Sanchez Telescope (TCS)}

We conducted an extensive observational program using the TCS, which is located at the Teide Observatory in the Canary Islands, Spain. This observatory is well-known for its good seeing (0.8--1.2 arcsec) and the high number of clear nights (about 80\%).

We collected data during 24 observational nights, spread between 3 October and 13 December, 2019. This time interval covers mainly the pre-perihelion passage of 2I/Borisov (which occurred on 8 December). The observing log is provided in Table~\ref{TCS2I_log}.
%The comet was observable at an airmass smaller than $\sim$2 for about 1-2 hours, before the morning twilight., thus we continuously acquired images during this time interval

\begin{figure}
    \centering
    \includegraphics[width=0.9\linewidth]{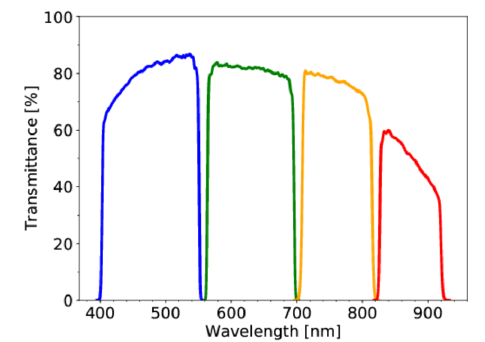}
    \caption{MuSCAT2 transmittance as a function of wavelength for g (blue), r (green), i (orange) and z$_s$ (red) bands \citep{2019JATIS...5a5001N}.}
    \label{fig:muscat}
\end{figure}

The TCS is a Cassegrain reflecting telescope with a 1.52 m diameter primary mirror and a focal length of f/13.8. For the MuSCAT2 instrument a system of lenses reduces the focal length to a ratio of $f/4.4$. This setup has four 1024 $\times$ 1024 pixel charge-coupled device (CCD) cameras with a scale of 0.434 arcsec per pixel. Each camera is designed to capture a specific band as follows. There are $g$ (400-550 nm), $r$ (550-700 nm), $i$ (700-820 nm), and $z_S$ (820-920 nm) cameras, providing simultaneous imaging in multiple bands using a dichroic mirrors system. The transmittance functions of these filters are illustrated in Fig. \ref{fig:muscat}~\citep{2019JATIS...5a5001N} and they also consider the quantum efficiencies of the CCDs.

The target was observable in the mornings, and the observations started when the object has risen above 27 $\deg$ altitude (because of the mount mechanical limit of the TCS telescope). Thus, we observed 2I/Borisov at an airmass of 1.5 -- 2. We acquired consecutive images with an exposure time of 30 seconds. The median number of images acquired during each session was 100.  All these were combined into a \emph{track and stack frame} (the images are shifted to follow the comet), and a \emph{sky frame} (the images are shifted according to sideral tacking). Because of the low airmass and telescope tracking issues, the seeing measured on the \emph{sky frame} ranged from 1.9 to 4.7 arcsec with a median value of 2.5 arcsec. 

\subsection{The 2.54-m Isaac Newton Telescope (INT)}

Spectro-photometric observations were also obtained with the 2.54-m INT telescope during five nights (Table~\ref{INT2I_log}), three at the beginning of October 2019 and two during the second half of November. This is located at the astronomical observatory Roque de los Muchachos Observatory, on the island of La Palma in the Canary Islands, Spain. The Wide Field Camera (WFC) was used. This instrument consists of four CCDs. To minimize the readout time a subframe of 1821 $\times$ 1821 pixels from CCD4 (the central one) was set for observations. This covers a field of view of 10.14 $\times$ 10.14 arcmin$^2$ at a sampling rate of 0.33 arcsec/pixel.

The images were acquired with the Johnson/Bessel filters $B$ and $V$, and with the Sloan filters $r$, $i$, and $z$. The following acquisition sequence was repeated during the observing time frame $B~-~V~-~R~-~V~-~i~-~V~-~z~-~V$. Because of the strong fringing effect in the $i$ and $z$ filters we discarded most of these observations.

Additionally, during the night of 25 September, 2019 we obtained a low resolution spectrum of 2I/Borisov using the Intermediate Dispersion Spectrograph (IDS)  with a configuration that includes the R150 grism and the EEV10 CCD camera. Because of the low signal to noise ratio of these data, the spectrum was only used to asses the overall spectral trend.

\subsection{The 4.2-m William Herschel Telescope (WHT)}

Spectroscopic observations of 2I were performed on the night of 26 November, 2019 (at 5.40 UT, 27 November) with the 4.2-m WHT located at Roque de los Muchachos Observatory, La Palma Canary Islands.

We used the Intermediate-dispersion Spectrograph and Imaging System (ISIS) mounted at the f/11 Cassegrain focus\footnote{\url{https://www.ing.iac.es/astronomy/instruments/isis/}}. In order to cover the spectral range from the ultraviolet to the near-infrared between 320 -- 900 nm, we used both the \emph{red} and the \emph{blue} arms.   For the \emph{blue} arm the setup included the R300B diffraction element with the default detector EEV12, which allows a dispersion of 0.86 \si{\angstrom}/pixel. In the same manner, for the \emph{red} arm is set with the R316R grating and the RED+ CCD camera, which provides a dispersion of 0.98 \si{\angstrom}/pixel. A slit with the width of 1 arcsec was put for both arms. With this configuration the resolution element for the \emph{blue} arm is 4.10 \si{\angstrom}, and for the red arm is 3.80 \si{\angstrom}.

The throughput of the dichroic (which allows to observe with both arms simultaneously) drops around 5500 $\pm$ 200 \si{\angstrom}. Also, for wavelengths longer than 8000 \si{\angstrom} the camera is affected by a strong fringing effect, while bellow 3500 \si{\angstrom} the signal for 2I/Borisov was buried in the noise. Consequently, we discarded the data from these intervals.

In order to remove the spectrum corresponding to Sun light, we observed with the same configuration three well known solar analogues, HD 30246, Hyades 64, and SA 98-978. The observations were made before (for the first two stars), and after (for SA 98-978) the observations of 2I, at similar airmasses. Additionally, to obtain the wavelength calibration the spectra of a CuAr + CuNe lamp were obtained before, between, and after the observations of the main target (three images).

Four exposures of 1200 seconds each were made for 2I. The data was acquired between 05:00 to 06:15 UT (the beginning of the first and last exposure), 27 November, 2019. During this time the airmass of the comet changed form 1.77 to 1.40. 

\section{Data reduction and analysis methods}

The first step for all data reduction (both spectro-photometric and spectroscopic) consists in applying the bias and  the flat field corrections. These calibration images were acquired at the beginning or at the end of each observing session (or observing night). Then, the spectro-photometric images were reduced using the Photometry Pipeline (PP) software (\citealt{Mommert_2017}), while the spectral data was reduced using Image Reduction and Analysis Facility - IRAF (\citealt{1986SPIE..627..733T}).

The analysis methods include the characterization of comet morphology, the computation of dust production rates using the model proposed by \cite{2012Icar..221..721F}, and the gas production rate using the Haser model \citep{Haser_1957}. For this analysis we used GNU Astro \citep{gnuastro, noisechisel_segment_2019}, Python \citep{10.5555/1593511} with astropy \citep{astropy} for ephemerides computations and other helpful operations, and PyMieScatt \citep{2018JQSRT.205..127S} for computing the scattering efficiencies at different refractive indices. Also, a part of the processing was done with GNU Octave (\citealt{octave}). 

\subsection{The spectro-photometric data}
We used the PP software (\citealt{Mommert_2017}) and several additional Python scripts in order to calibrate the images. The PP software is an envelope for SExtractor (Source Extractor) and SCAMP \citep{2006ASPC..351..112B}. The astrometric and photometric calibration are performed using the large data catalogues generated by GAIA \citep{2018A&A...616A...1G} and PanSTARRS \citep{2016arXiv161205560C} surveys, respectively. 

%The astrometric calibration is performed in order to identify the positions of background stars. Afterwards, PP offers a technique for adjusting the instrumental magnitude to match the magnitudes derived from PanSTARRS survey data. The process involves evaluating various background objects against catalog data to ascertain the zeropoint. The outcome of this photometric calibration function is the zeropoint value and its associated uncertainties for every image.
For the data obtained with TCS we removed the camera patterns uncorrected by the bias and flat procedures using the \emph{astnoisechisel} routine (from GNU Astronomy package ) with the default parameters \citep{gnuastro, noisechisel_segment_2019}. The \emph{astnoisechisel} is a noise-based non-parametric technique dedicated to the detection of very faint nebulous objects which are buried in noise. This tool detects the sky background as an intermediary step, and allows to remove it. We also used the \emph{astnoisechisel} procedure to search for the faintest structures of the 2I/Borisov coma.

The astrometric calibration is performed for every image.  Then, for each observing session and for each band, the images are combined in order to obtain the \emph{track and stack frame} -- by shifting the images to follow the apparent motion of 2I/Borisov at that time, and the \emph{sky frame}. For combining the images we used two methods, the median and the average. The median has the advantage of removing most of the background stars from the track and stack frame, but it may introduce uncertainties in the photometric data (this was resolved by comparing the results with those obtained by combining the images with the average method). The photometric calibration constant, the zero point, was determined using the combined sky frame. The uncertainty of these determinations depends on the number of stars available for each frame and it spans between 0.02 -- 0.06 mag.

\subsection{The spectral data}
The spectroscopic observations were reduced with the help of IRAF, specifically the \emph{apall} routine, which is included in \emph{noao.twodspec.apextract} package. The spectrum of the comet was extracted using an aperture of 4 arcsec, while the sky background was measured between 10 to 20 arcsec from the central trace, on each side of the spectrum.  Each spectral image was reduced in the interactive mode of IRAF, which allowed us to chose the optimal method for tracing the spectrum.

After the spectra extraction from the images, we processed the results using GNU Octave \citep{octave}. We made the wavelength calibration using the lines of the CuAr + CuNe lamp. We corrected for the atmospheric extinction using the model described in the technical notes provided by the Isaac Newton Group\footnote{\url{https://www.ing.iac.es//Astronomy/observing/manuals/ps/tech_notes/tn031.pdf}}. Then, we divided the obtained spectrum of 2I with that of the analogue solar Hyades 64 as shown by \cite{2022A&A...664A.107T}, this star being one of the best solar analogues match across the 350 - 800 nm. The other two stars were used for comparison. The flux calibration was performed considering the apparent magnitude obtained in the $g$ filter with TCS which was observing simultaneously. According to \cite{Willmer_2018}, the spectral density flux zeropoint of Vega for $g$ band is 20.64, corresponding to a value of \SI{5.52e-9}{\erg\per\second\per\cm\squared\per\angstrom}. Using this value and integrating the spectral density flux between 400 and 550 nm, we obtained the normalization coefficient in order to re-scale the spectrum density values to the absolute ones.

\subsection{Analysis methods}
{\bf Dust model.} The dust model, as detailed in Appendix~\ref{apx:dust}, follows the approach described in \cite{2012Icar..221..721F} to determine the dust production rate of comet 2I/Borisov. The model incorporates assumptions of constant outflow velocity for particles of the same radius, $a$, and spherically symmetric dust distribution. The dust production rate, $Q$, is derived considering the terminal velocity of dust particles as estimated from the values of the coma length scales in the sunward direction. 
The model accounts for the size distribution of particles using the distribution proposed by \cite{2012Icar..221..721F} (UF distribution),
\begin{equation}
    \frac{dQ}{da} = g_0 e^{-\frac{a_0}{a}}\Big(\frac{a_0}{a}\Big)^\alpha
\label{eq:dust_dist}
\end{equation}
 It allows for the computation of total dust grains, mass, and dust production rate. The parameters in the model include a normalization constant, denoted as $g_0$, and two free constants, namely $a_0$ and $\alpha$. The size distribution's peak, indicating the most probable size, corresponds to $a_p$, which is calculated as the ratio of $a_0$ and $\alpha$.

{\bf Haser model.} For the analysis of the molecular emission from comet 2I/Borisov, the Haser model, outlined in Appendix~\ref{apx:haser}, is employed. This model extrapolates the total number of molecules in the coma by considering a radial distribution based on the parent and daughter scale lengths. The fluorescence efficiency, or "g-factor", is used to compute the total number of molecules in the entire coma. The production rate of molecules is then derived from measured fluxes of specific molecular emission bands, ensuring static equilibrium between destruction and production rates. 

\section{Results}
In this section we present both the photometric and the spectroscopic analysis of the observational data obtained with TCS, INT and WHT. The TCS/MuSCAT2 observations of the comet allows us to characterize the comet's morphology and its changes over 3 October - 13 December, 2019 time frame. The results are presented in the first subsection.
%, where we employed only the data gathered with TCS for consistency covering a longer time frame. 
The dust colors and its production rate are described in the second and the third subsections, where we also cross-check the TCS/MuSCAT2 values with those obtained with the INT/WFC instrument. This provides a strong reliability of our determinations (both telescopes observed simultaneously during the night of 4 and 5 October, 2019). The last subsection describes the spectroscopic results obtained with WHT on the morning of 27 November, 2019.

\subsection{Morphology}

The comet 2I/Borisov exhibits a complex structure as a result of its interaction with the Sun. The radiation pressure is influencing the shape of the coma and how the tail is oriented. As a first step, we analyze the morphology using the TCS data in order to describe quantitatively how the solar radiation is affecting the coma. Then, the large number of observations allows to make estimations on the terminal speed of the coma dust particles. Another key question would be if there are any color variations along the brightness profile of the comet. Studying the comet's morphology in four different bands can provide answers to this.

\begin{figure}
    \centering
    \includegraphics[width=0.9\linewidth]{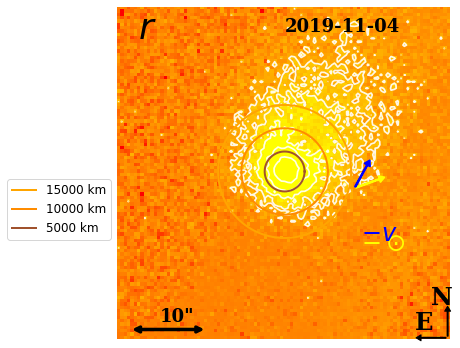}
    \caption{Composite image in r-band with the comet on November 4, 2019 obtained with 98 co-added frames. We illustrate the north (N) and east (E) directions of the equatorial coordinate system, together with the image scale of 10 arcseconds. The anti-solar (\textbf{-$\mathbf{\odot}$}) and anti-velocity (\textbf{-v}) are plotted with yellow, and blue, respectively. The three circles show the corresponding apertures of radii 5000 km, 10\,000 km and 15\,000 km. }
    \label{fig:comet}
\end{figure}

An $r$ band stacked image is shown in Figure \ref{fig:comet}, with contours added (using the \emph{matplotlib} library from Python) to distinguish between different intensity levels. The coma appears bright and the particle density decreases with distance from the nucleus. In addition, the anti-velocity and the anti-solar vectors are plotted to illustrate the orientation of the coma, which is developing on the north side with respect to the equatorial coordinate system, under the influence of both inertial and non-gravitational forces.

\begin{figure}
    \centering
    \includegraphics[width=0.9\linewidth]{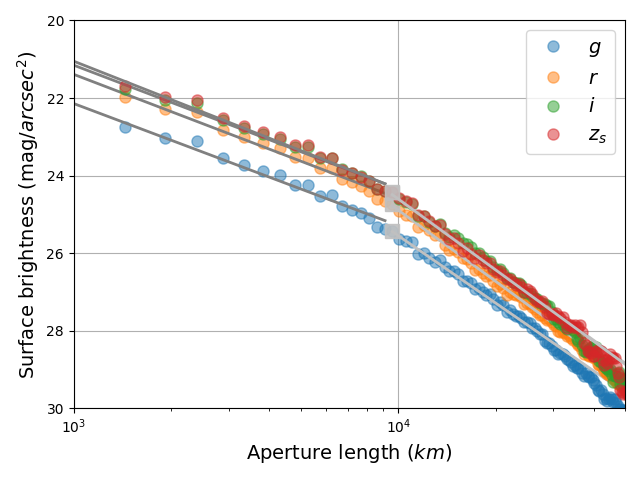}
    \caption{Surface brightness profile of 2I plotted for each spectral band on 4 November 2019. For each band, the profile is fitted by two grey lines, while the square marks the beginning of the distortion due to the radiation pressure effects.}
    \label{fig:profile}
\end{figure}

A clear coma resembling that of a comet followed by a dust tail can be observed from the processed images. We retrieved the comet profile by computing the surface brightness, defined as the brightness per unit angular area, $\Sigma_\lambda = \frac{dB_\lambda}{d\Omega}$. We estimate it by taking consecutive circular apertures with a step of 200 km. The flux is interpolated using the \emph{aperture\_photometry} of \emph{photutils} with the \emph{subpixel} method, dividing each pixel in 25 subpixels. As one can see in Fig. \ref{fig:profile}, there are no discernible differences in morphology between the surface brightness profiles retrieved from the images taken simultaneously in the four spectral bands.

For each profile, we estimate the coma length scale, $X_R$, by spotting the point of coma distortion (the "knee"), where the profile trend is subject to an obvious change in steepness \citep{1991ASSL..167...19J}. To find this point we apply a linear regression algorithm fitting the profile using two lines. The slope of each line gives us the surface brightness gradients and with the "knee" position we determine the coma length scale. 

The length scale is averaged over values estimated for the four bands, and plotted with respect to the square of the heliocentric distance (see Fig. \ref{fig:x2fit}) expressed in squared astronomical units, $au^2$. As we expect a proportional relationship (see Appendix~\ref{apx:dust}), we perform a linear fit fixing the intercept at the origin. We obtain that $X_R/r^2=2000 \pm 300$ km/au$^2$ leading to 
\begin{equation}
\label{eq:speed}
 v_d(a) = (0.15 \pm 0.01) \dfrac{m}{s} \Big(\dfrac{a}{\SI{1}{\meter}}\Big)^{-0.5}
\end{equation}
The Eq.~\ref{eq:speed} describing the dust particles' terminal speed with respect to their size is in agreement with the Monte Carlo simulations result of $0.1423a^{-0.5}$ \citep{2011ApJ...732..104T}, where the magnitude of the velocity is determined in $m/s$ if we express $a$ in meters.

The surface brightness gradients monitored during all the observing sessions -- the slopes of the two linear fits, $m=\frac{d \log \Sigma_\lambda}{d\log \theta}$, show us that for the coma we get values between $m=-1.2$ and $m=-1.5$, having an increasing trend near perihelion. This transition from an almost isotropic coma, characterized by $m=-1$ \citep{1991ASSL..167...19J}, is the effect of the solar radiation pressure. Beyond the "knee", the steepness drops to values around $m=-2.5$.

\begin{figure}
    \centering
    \includegraphics[width=0.9\linewidth]{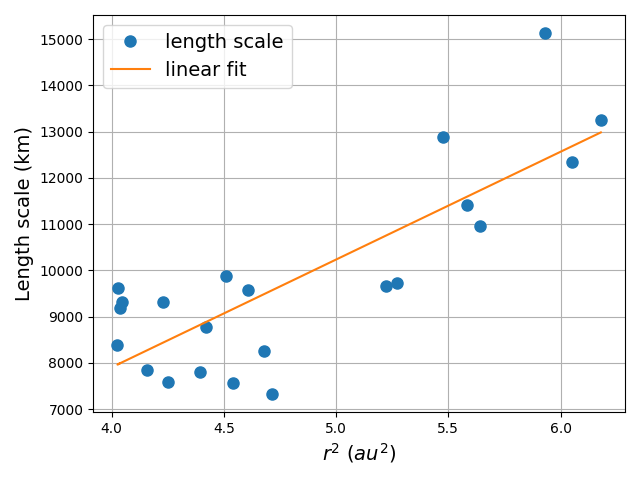}
    \caption{The coma length scale as a function of the squared heliocentric distance. Data is fitted linearly with fixed intercept: $y=2000x+0$.}
    \label{fig:x2fit}
\end{figure}

\subsection{Dust colors}

We calculate the flux collected inside an apparent aperture size of 15,000 km for both TCS and INT. This aperture corresponds to an angular size ranging from 8.3 to 10.3 arcsec , depending on the Earth - comet distance at the observing time. From the photometry of TCS data, we find the average color indices over all the observed nights and the corresponding standard deviations, $(g-r)=0.61 \pm 0.08$, $(r-i)=0.16 \pm 0.07$, and $(i-z_s)=0.10 \pm 0.06$. On the other hand, with INT we find $(r-i)=0.22 \pm 0.04$ and $(B-
V)=0.79 \pm 0.04$.

\begin{figure*}
    \centering
    \includegraphics[width=0.9\linewidth]{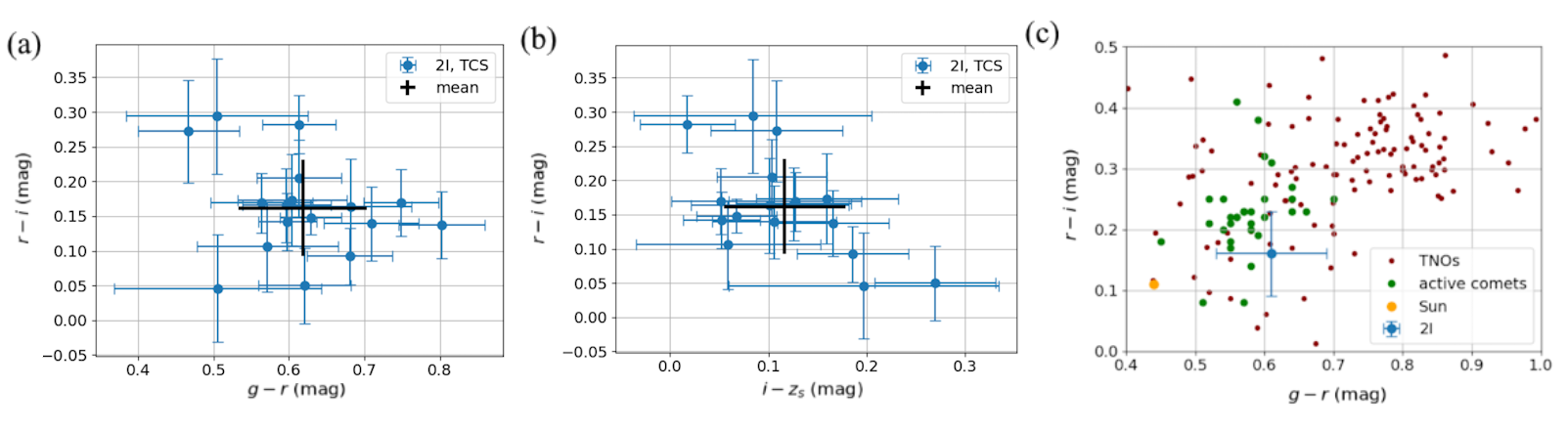}
    \caption{Color diagrams of 2I observed with TCS data: (a, b) $(r - i)$ vs. $(g - r)$, and $(i-z_s)$ colors, respectively, plotted with errorbars for each night of observations, the mean value is shown with orange; (c) $(r - i)$ vs. $(g - r)$ of 2I plotted compared with the colors of other active comets from the Solar System \citep{Solontoi_2012} and the trans-Neptunian objects -- TNOs \citep{2012ApJ...749...10O}. Also, the Sun color is plotted to emphasize the reddening. }
    \label{fig:colors}
\end{figure*}

The first two plots of Fig. \ref{fig:colors} show the color indices retrieved from the TCS measurements. They seem to clusterize around the mean value exhibiting several variations within the time frame of the observational campaign. Most of these variations are probably the result of the observational errors. The larger offsets corresponding to the top-left side of the plots correspond to measurements at different dates, not being correlated to a specific time interval. Overall, the colors align closely with the characteristics of Solar System long-period \citep{Solontoi_2012} and Trans-Neptunian objects \citep{2012ApJ...749...10O} as shown in the last plot of Fig. \ref{fig:colors}. 

As it can be seen in Table \ref{table:colors_comparison}, our findings are in agreement with the results reported by other groups. Overall, 2I does not seem to suffer obvious compositional changes that may lead to a deviation in color indices.

\begin{figure*}
    \centering
    \includegraphics[width=0.9\linewidth]{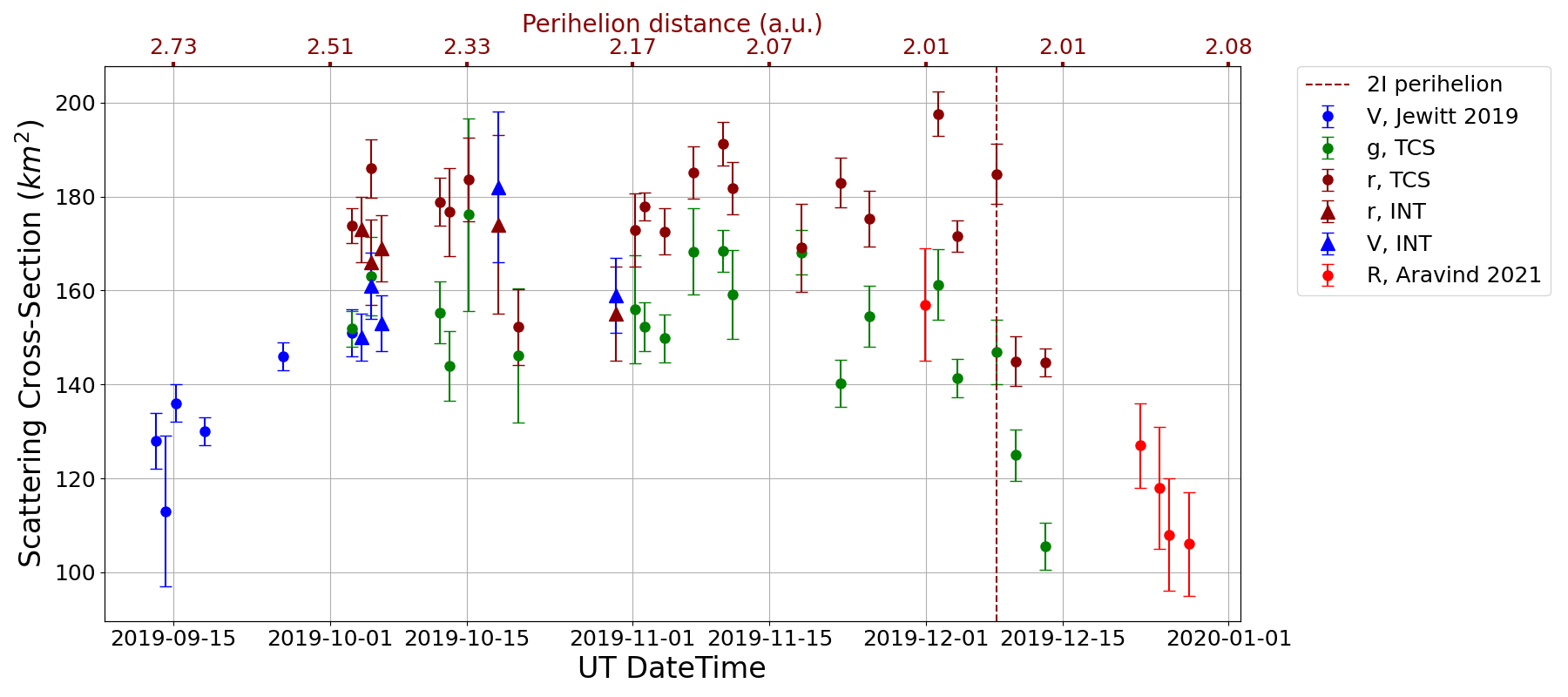}
    \caption{The effective cross-sections in $g$ and $r$ for TCS, and $V$ and $r$ for INT are represented with respect to time. All the values are determined within a fixed circular aperture radius of 15,000 km. We plot for comparison the values reported by \citep{Jewitt_2019} in $V$ band (15,000 km aperture) and \citep{Aravind_2021} in $R$ band (10,000 km aperture).}
    \label{fig:cs}
\end{figure*}

\begin{table*}
\centering
\begin{tabular}{cccccc}
           & Time Interval & $g-r$              & $r-i$              & $i-z_s$  & $B-V$         \\ \hline
TCS &   Oct 3 - Dec 13         &  $0.61 \pm 0.08 $ & $0.16 \pm 0.07 $ & $0.10 \pm 0.06 $ &-\\
INT &   Oct 4 - Oct 30        &  - & $0.22 \pm 0.04 $ & - & $0.79 \pm 0.04$\\
\cite{Guzik_2019}      & Sep 10-13            &    $0.63 \pm 0.02$  &  - & -&-\\
\cite{2020MNRAS.495.2053D}      & Sep 13-26            &    $0.63 \pm 0.02$  &  - &-&- \\
\cite{Hui_2020} & Sep 26 - Jan 26               &     $0.68 \pm 0.04$  & $0.23 \pm 0.03$  &  - &-\\ 
\cite{Bolin_2020}      & Oct 12            &   $0.63 \pm 0.05$  & $0.20 \pm 0.05$  &  -&- \\
    & Sep 12            &   -  & -  &  -& $0.76 \pm 0.12$\\
 \cite{opitom_2019}   & Sep 30 - Oct 20            &   -  & -  &  -& $0.82 \pm 0.02$\\
 \cite{Aravind_2021}   & Nov 30 - Dec 27           &   -  & -  &  -& $0.80 \pm 0.05$\\

\hline
The Sun \citep{Willmer_2018}       & & $0.44 \pm 0.02$  & $0.11 \pm 0.02$  & $0.03 \pm 0.02$ & $0.64 \pm 0.02$
\end{tabular}
\caption{Average colors of 2I/Borisov during TCS and INT campaigns and their uncertainties ($\pm \sigma$). Other values retrieved from the literature and the solar colors are provided for comparison. }
\label{table:colors_comparison}
\end{table*}

\begin{figure*}
    \centering
    \includegraphics[width=0.9\linewidth]{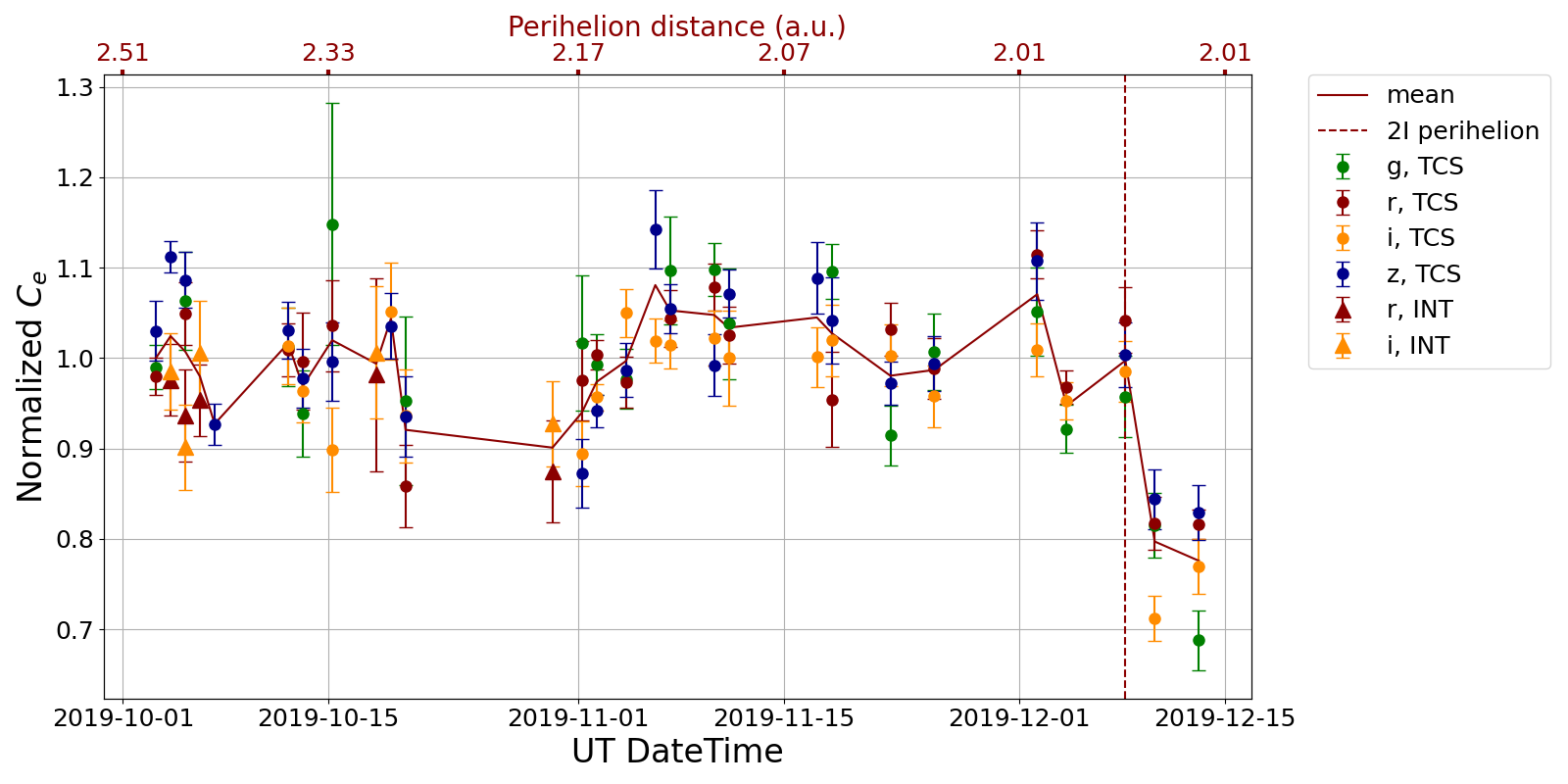}
    \caption{The effective scattering cross-section with respect to time, and normalized to the median value along each band. Both TCS and INT data are plotted. The dark red line connects the mean values corresponding to each night. The perihelion is marked with a dashed line.}
    \label{fig:ncs}
\end{figure*}

\subsection{Dust production and mass loss}

The absolute magnitude and the effective scattering cross-section allow to characterize the comet activity and subsequently, to calculate the dust production rate and the mass loss. Eq.~\ref{ce} provides the formula for $C_e$ -- the effective scattering cross-section, where $r_\oplus=\SI{1.5e8}{\kilo\meter}=1$ au is the Earth-Sun mean distance, $p_\nu$ is the geometric albedo taken as $0.1$ \citep{ZUBKO2017104} and $m_{\odot,\nu}$ is the apparent magnitude of the Sun in the band $\nu$ $\in \{g,r,i,z_s\}$, and $H_\nu$ is provided by the Eq.~\ref{Hnu}. When estimating the flux and computing these quantities, we always take an aperture radius of 15\,000 km. The phase term is estimated linearly \citep{1991ASSL..167...19J} as $2.5\log \phi(\alpha)\approx -0.04\alpha$.

The photometry allows us to compute the effective scattering cross-section in each band,
\begin{equation}
    C_e = \frac{\pi r_\oplus^2}{p_\nu}10^{-0.4(H_\nu-m_{\odot, \nu})}
\label{ce}
\end{equation}

\begin{equation}
H_\nu=m_\nu - 5\log (r/r_\oplus) - 5\log(\Delta/r_\oplus) + 2.5\log \phi(\alpha)
\label{Hnu}
\end{equation}

Our results obtained for the effective cross-sections in $g$ and $r$ bands are compared in  Fig. \ref{fig:cs} with the results found by \cite{Jewitt_2019} for V-band (aperture radius of 15\,000 km)  and \cite{Aravind_2021} in R-band (aperture radius of 10\,000 km). Our data fills the gap in agreement with the measurements reported by the other authors. 

In Fig. \ref{fig:ncs}, the cross-sections of each band are normalized by dividing them to the median value. One can notice a trend consistency along the bands. There are small increases at the beginning of each month that may suggest a periodical behaviour during October - December 2019, or they are simply fluctuations and the periodic pattern is coincidental. In general, the cross-section experiences fluctuations around an average value ($\SI{155}{\kilo\meter\squared}$ in $g$,
$\SI{178}{\kilo\meter\squared}$ in $r$, $\SI{188}{\kilo\meter\squared}$ in $i$, $\SI{206}{\kilo\meter\squared}$ in $z_s$) at distances between 2.51 au to 2.01 au, before the perihelion passage. These variations are barely exceeding 10\% with respect to the median $C_e$ for each band. However, they are consistent and large enough ($\approx2\sigma$) to not be considered errors. This behaviour persists until December 3, where a noticeable declining trend becomes evident following the last peak. 

Applying the dust model, as detailed in the methods section (Appendix~\ref{apx:dust}), we aim to estimate the dust production rate and total mass loss. The normalization constant $g_0$ is determined through numerical integration of Eq. \ref{eq:Ce_dust_dist}, according to the effective scattering cross-sections' estimated values. Our analysis, illustrated in Fig. \ref{fig:colors}, reveals an increasing spectrum intensity towards redder colors. This trend signifies an intensification of scattering, resulting in larger values for $C_e$. These observations aligns with our measurements, where the dependence $C_e(\lambda)$ demonstrates a roughly linear increasing trend. Thus, it is imperative that our model accurately reproduces this linear trend, and achieving this requires an appropriate size distribution of dust particles having a specific scattering behaviour. As mentioned in Appendix~\ref{apx:dust}, the scattering coefficients depend on the refractive index of the material ($m=n+ik$). It is very much likely that the dust coma is a mixture of particles made of distinct materials decoupled from different regions of the surface layer of the comet nucleus.  

\begin{figure}
    \centering
    \includegraphics[width=0.9\linewidth]{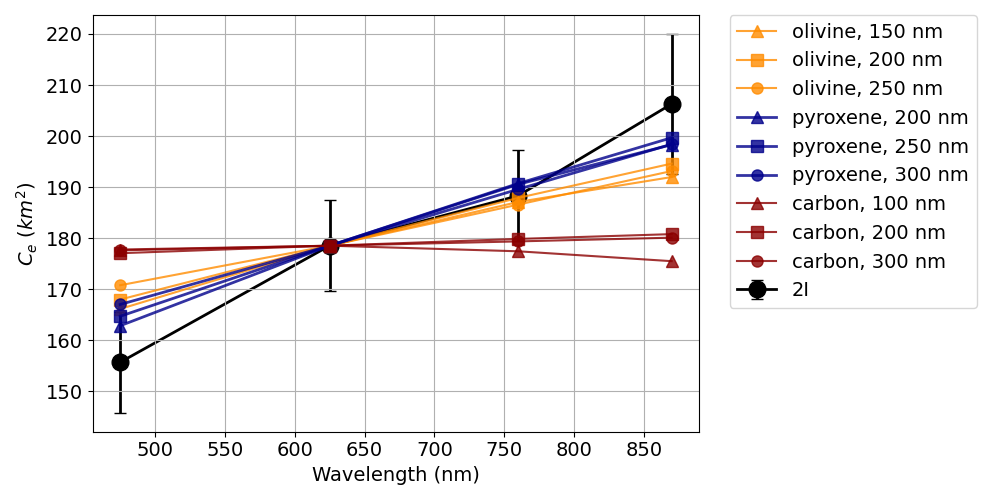}
    \caption{Average effective scattering cross-section of 2I before perihelion with respect to time, compared to the same cross-section obtained via Mie scattering simulations. Different dust particle distributions and different materials are employed to replicate the observational data results (olivine - yellow lines, pyroxene - blue lines, carbon - red lines). The most probable particle size, $a_p$, for each distribution is mentioned in the legend.}
    \label{fig:cs_sim}
\end{figure}

Through multiple simulations involving various materials, such as olivine ($m=1.80+0.10i$), pyroxene ($m=1.70+0.04i$), and carbon ($m=2.00 + 0.50i$), we have identified that particles with the most probable size of around 200 -- 300 nm emerge as the primary scatterers in the coma (15\,000 km aperture). In Fig. \ref{fig:cs_sim} is shown a summary of the resulting curves, which were obtained using the UF distribution with $\alpha=4.7$, a value found via grid searching. 

We can observe that pyroxene with $a_p=250$ nm would be the best choice when reproducing the scatterers behaviour of 2I. Olivine would be compatible as well. However, carbon does not seem to be predominant in the composition of 2I, for which we assume that could be a mixture of olivine and pyroxene, or another mixture of materials with similar scattering properties, i.e. refractive indices.

 Let us consider the density of compact dust as 2650 kg/m$^3$ \citep{Yang_2021} and the particle size distribution found for pyroxene with $a_p=250$ nm. Thus, the mean value of the effective cross-section before perihelion leads to an estimation of the total mass of dust equal to $(1.9 \pm 0.2) \times 10^6$ kg and a dust production of $(7.0\pm0.4) \times 10^{14}$ particles/s, equivalent to $(4.0\pm0.2)$ kg/s. After the perihelion passage, based on the same considerations, the total mass of dust decreases with $(0.6 \pm 0.2)$ kg/s assuming a linear trend. The dust production is decreasing as well after perihelion, reaching a value of 3.3 kg/s on 13rd December 2019.

\subsection{Spectroscopy}

\begin{figure*}
    \centering
    \includegraphics[width=0.9\textwidth]{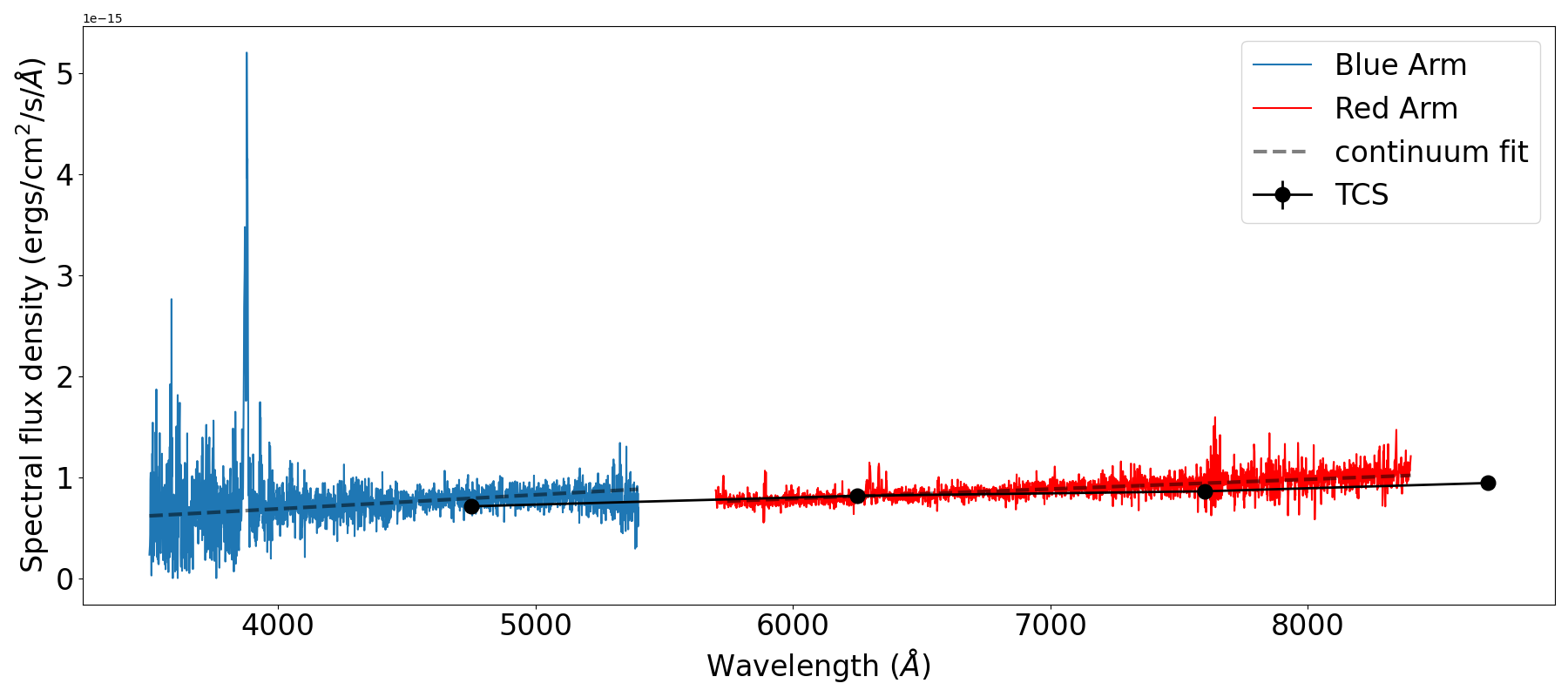}
    \includegraphics[width=0.9\textwidth]{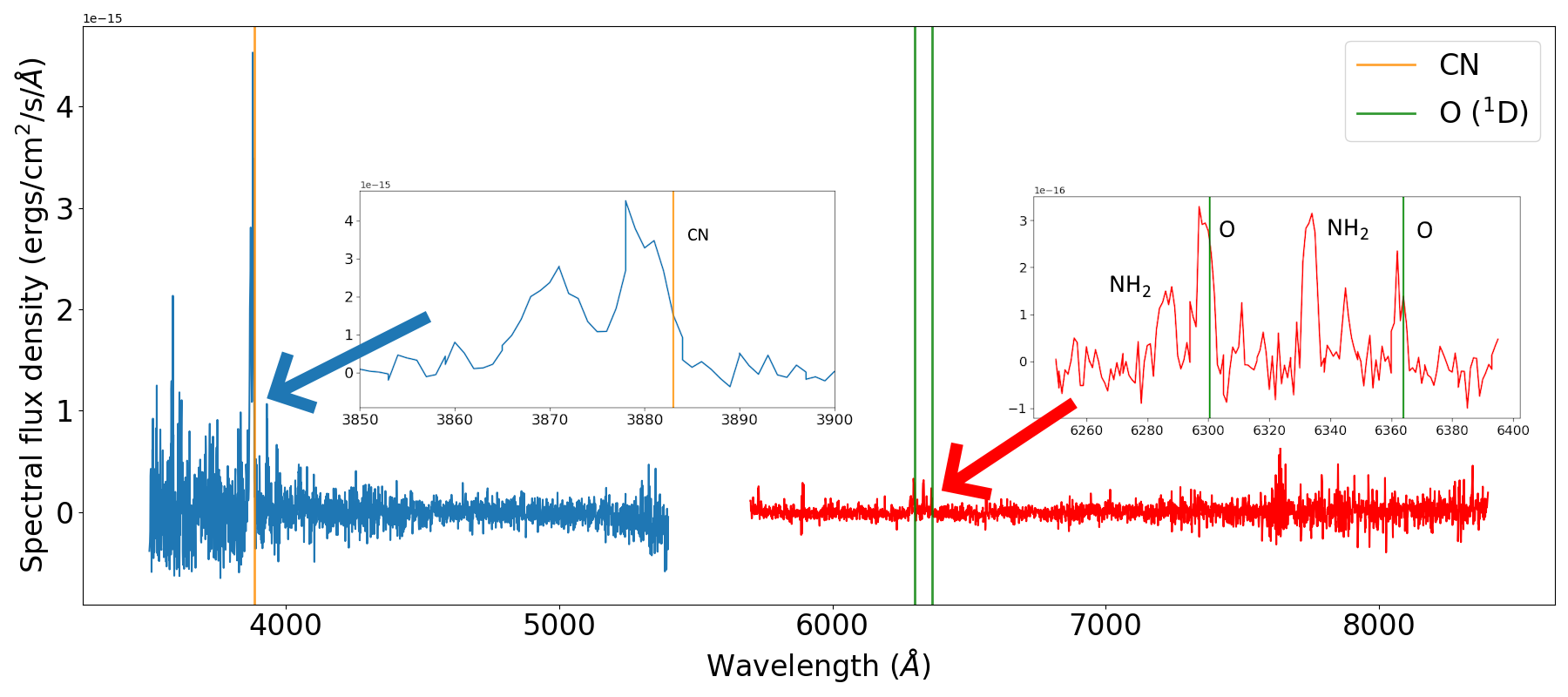}
    \caption{The comet spectrum obtained with WHT on 2019 November 26. The upper spectrum contains the dust continuum, whereas the bottom one is the fluorescence spectrum with CN, NH$_2$, and O I bands highlighted.}
    \label{fig:spectra}
\end{figure*}

The spectrum of 2I/Borisov obtained with WHT is shown in Fig. \ref{fig:spectra}. We merged the data obtained with both arms of the spectrograph. We show the comparison between the spectral data and the corresponding values of the flux obtained with TCS at the effective wavelengths of each band ($g$, $r$, $i$, $z_s$).

To calibrate the spectrum we used the apparent magnitudes obtained with TCS on the morning of November 25 (two days prior to the spectral observations). Thus, the method implies an uncertainty of $\approx$ 0.05 mag caused by the inaccuracy of the photometric determination, and by the different observing dates. First, we applied the transfer function of the $g$ filter (modulated by the quantum efficiency of the camera) to the spectral data. The total response is obtained in ADU (analog to digital units). Then, the calibration consists in determining  the spectral flux density values using the zeropoint for Vega \citep{Willmer_2018} and the corresponding apparent magnitude in $g$-band obtained with TCS. 
%We analyse the spectral data obtained with WHT on  the morning of November 27. We processed both B and R filters subtracting the solar spectrum. The reduced spectrum is presented in the upper plot of .

To characterize the dust continuum the spectrum is fitted linearly. We find $S'_B=\SI[separate-uncertainty = true]{18 \pm 2}{\%\per1000\angstrom}$ in the range $4000-5000 \ \si{\angstrom}$ and $S'_R=\SI[separate-uncertainty = true]{11 \pm 1}{\%\per1000\angstrom}$ in the range $5700-7250 \ \si{\angstrom}$. The value obtained in the blue domain is consistent with those reported by \cite{Fitzsimmons_2019} for September, $\SI[separate-uncertainty = true]{19.9}{\%\per1000\angstrom}$, or \cite{Lin_2020} for the beginning of November, $\SI[separate-uncertainty = true]{19.3}{\%\per1000\angstrom}$. Further, the red gradient shows that the spectrum of 2I/Borisov is steeper along the blue domain as suggested by \cite{2020MNRAS.495.2053D} as well.

To detect the emission bands, we subtracted the underlying continuum caused by sunlight reflected by dust. The comparison of the spectrum before and after removing the dust continuum is depicted in Fig. \ref{fig:spectra}, revealing the prominent CN (0-0) gas emission attributed to solar fluorescence at 388 nm.  The comet spectra showcasing solely gas fluorescence emission characteristics is provided in the bottom plot of Fig. \ref{fig:spectra}, where we spotted also the red doublet of oxygen at 630 nm and 636 nm. 

We integrated numerically the total flux inside the CN band obtaining $(4.8 \pm 0.2) \times 10^{-14}$ ergs/cm$^2$/s. To estimate the production rate, we use the Haser model (see Appendix~\ref{apx:haser}). An implementation of this model is provided by the Lowell Minor Planet Services through their Comets API \footnote{\url{https://asteroid.lowell.edu/comet/ghq}}. Taking into account the ephemerides of 2I on 27th November, $r=2.03$ au and $\dot{r}=-5.2$ km/s, the interpolated g-factor is $3.6\times 10^{-13}$ ergs/s/molecule \citep{2010AJ....140..973S}. The parent and daughter molecule lengthscales are those established in \cite{1995Icar..118..223A}. Thus, we find $Q(CN)=(1.2\pm0.1)\times 10^{24} s^{-1}$, being close to the value of $(1.5\pm0.5)\times 10^{24} s^{-1}$ found by \cite{bannister2020interstellar} on the same day, November 26, or $(1.6\pm0.5)\times 10^{24} s^{-1}$ reported by \cite{opitom_2019} on October 20. It seems that the production rate of CN follows a decreasing trend as \cite{Fitzsimmons_2019} report $Q(CN)=(3.7\pm0.4)\times 10^{24} s^{-1}$ on September 20. Moreover, another measurement in late September 2019 leads to $Q(CN)=(2.3\pm0.4)\times 10^{24} s^{-1}$ \citep{2020MNRAS.495.2053D} and \cite{opitom_2019} find values between $(1.8 - 2.1)\times 10^{24} s^{-1}$ for the first half of the next month. Post-perihelion, \cite{Cordiner_2020} report production rates of $(0.9-1.2)\times10^{24} s^{-1}$ for the timeframe December 15-19.

In our spectroscopic analysis, we successfully identify the OI red doublet lines at 630 nm and 636 nm. However, we note that the green line of the forbidden oxygen lines (out at approximately 557.7 nm) is outside the instrument range, preventing its detection in our observations. Also, the resolution is not sufficient to separate the red OI cometary lines from the telluric ones. The two lines are partially overlapping each other as it can be seen in Fig.~\ref{fig:spectra}. The forbidden oxygen lines were detected by \cite{opitom_2021_b,McKay_2020} as well. 
We observe two lines at 629 nm and 633 nm corresponding to ammonia, NH$_2$. Assuming the g-factor value of $0.564 \cdot 10^{-3}$ photons/s determined by \cite{kawakita_2001} for the $\Tilde{A}$(0, 8, 0) $\Pi$ band of NH$_2$ at 629 nm \citep{1988PASP..100..702P}, we derive $Q(NH_2)=(0.7 \pm 0.2) \times 10^{24} s^{-1}$, meaning that the gas production quotient between NH$_2$ and CN is $\log(NH_2/CN)=-0.2\pm0.2. $The signal-to-noise ratio for this measurement is $S/N \approx 7$. If we consider the lower values of the fluorescence efficiencies for other lines that are in the spectral interval covered by WHT, we expect their signal to be undetectable at this noise level. This is confirmed by our analysis. On the other hand, the signal detected at 633 nm is actually an overlapping of two unresolved NH$_2$ bands \citep{Fink_1990, Fink_1996, Arpigny_1987} and from the total detected flux we obtain an upper limit for these bands of $(1.1 \pm 0.2) \times 10^{24} s^{-1}$.  
%  \cite{10.1093/mnras/stab2609} have reported a value of $\log(NH_2/CN)=0.8\pm0.1$ for the comet 8P/Tuttle at 1.2 au. 
Regarding C$_2$, we do not detect any lines above the noise level.

\section{Discussion and conclusions}
In this study, we presented a comprehensive analysis of observational data derived from a monitoring campaign focused on the first interstellar comet, 2I/Borisov.

By using the observations obtained with TCS, INT and WHT telescopes, we offer key insights into the comet's morphology and spectro-photometry:
\begin{itemize}
    \item 2I/Borisov exhibited clear signs of activity, featuring a coma length scale extending between 8000 and 13000 km. The observed length showed a linear trend with respect to the squared heliocentric distance.
    \item Simultaneous imaging in four bands revealed a uniform composition across the entire comet surface, indicating homogeneity in morphology. Color indices indicated a reddish hue, akin to active comets within our solar system and Trans-Neptunian objects.
    \item The comet's profile indicated a diffuse nature, aligning with typical characteristics of Solar System comets. The variation in surface brightness gradient observed in our study ranged from $m~=~-1.2$ to $m~=~-1.5$, which, appeared less steep than the gradient reported by \cite{Jewitt_2019} for September 26 ($m=-1.8$). The discrepancy can be attributed to the changes occurring in the comet's behavior, particularly around the onset of October, when our analysis identified an increasing trend in activity. This minor outburst is followed by a stationary trend of the cometary activity exhibiting small fluctuations. This persists until December. The first jump corresponding to October has been also spotted by \cite{Bolin_2020}, where it was proposed that the comet passed the water-ice line at 2.5 au as a potential cause. 
   \item The assumption of small fluctuations in the comet activity trend prior to perihelion is based on the presence of two more activity jumps in early November and December as suggested by the trend of the effective scattering cross-section. 
   Regarding the magnitude of the cross-section, \cite{Kim_2020} found in the same timeframe an average value of roughly $200$ km$^2$ for the effective scattering cross-section in V-band for an aperture of 16000 km, whereas we found the values of $(155 \pm 9)$ km$^2$ in $g$ band and $(178 \pm 8)$ km$^2$ in $r$ band for an aperture of 15000 km radius.
   \item Employing the acquired data, we computed the dust production rate and mass loss by establishing a dust coma of particles made from pyroxene with the most probable size equal to 250 nm. Using the distribution proposed by \cite{2012Icar..221..721F} we computed a dust production rate of roughly $ 7 \times 10^{14}$ particles/s before perihelion, equivalent to 4 kg/s assuming $\rho=2650$ kg/m$^3$. For comparison, \cite{Clements_2021} determined a dust production between 11 -- 16 kg/s (assuming large dust particles with $\overline{a}=100 \ \mu m$ and $\rho=1000 \ kg/m^3$, and slow ejection speeds between 5 -- 9 m/s),  exhibiting a slightly decreasing trend after perihelion. Based on the same model and assumptions, similar trend and values were obtained by \cite{Epifani_2021}. Depending on the ejection speed value, they find a dust production rate between 3 -- 25 kg/s in October, and 2 -- 18 kg/s in December.
   
   \cite{Manzini_2020} argued that the emitted dust, once released, disperses over a relatively short distance from the nucleus. Subsequently, it rapidly undergoes deflection under the influence of solar radiation pressure, eventually amalgamating into the comet's tail implying a low ejection speed. They found particles of 30 -- 100 $\mu$m being predominant in the tail, expecting relatively smaller particles closer to the nucleus, in agreement with the simulations performed by \cite{2020MNRAS.495.2053D} and \cite{Kim_2020}. Our analysis was focused on the dust emissions within a 15000 km aperture, i.e. the coma dust. We found that the most probable size of dust particles in the coma is in the range 200 -- 250 nm, much smaller than in the tail. According to Eq.~\ref{eq:speed}, the ejection speed would be around 300 m/s, high enough to escape from the comet's tail.
   \item The activity began diminishing in December 2019, after the perihelion passage. This behaviour was reported by other groups as well \citep{Kim_2020, Aravind_2021, Hui_2020} and it is quantified by estimating linearly the decrease rate of the effective scattering cross-section, $dC_e/dt$. We found that $dC_e/dt=(-4.4\pm1.6) \ km^2/d$ during a time span between 2 December and 12 December 2019, inside a 15000 km aperture radius. Subsequently, we determined the net mass loss of dust in the coma of approximately --0.6 kg/s during this post-perihelion interval. Considering a longer period of time after the perihelion passage, \cite{Aravind_2021} found $(-1.77 \pm 0.22) \ km^2/d$ for an R-band aperture radius of 10000 km between 30 November and 27 December 2019 leading to a net mass loss of --2.7 kg/s assuming a mean size of dust particles equal to $\overline{a}=100 \ \mu m$ and $\rho=1000 \ kg/m^3$. On the other hand, \cite{Hui_2020} reported a decrease rate $dC_e/dt = -0.43 \pm 0.02 \ km^2/d$ for an r-band aperture of same size, 10000 km, deriving a net mass loss of $-0.4$ kg/s, assuming $\rho=500 \ kg/m^3$. The discrepancy in the values of $dC_e/dt$ is attributed to the different observation timeframes, with \cite{Hui_2020}'s observational campaign extending until late January. In addition, our result is derived using a 15000 km aperture radius, a larger value being expected from this point of view.
   
\end{itemize}

\section*{Acknowledgements}

We acknowledge financial support from the Agencia Estatal de Investigaci\'on of the Ministerio de Ciencia e Innovaci\'on MCIN/AEI/10.13039/501100011033 and the ERDF “A way of making Europe” through project PID2021-125627OB-C32, and from the Centre of Excellence “Severo Ochoa” award to the Instituto de Astrofisica de Canarias.
This work is partly supported by JSPS KAKENHI Grant Number JP18H05439, JST CREST Grant Number JPMJCR1761. This paper is based on observations made with the MuSCAT2 instrument, developed by ABC, at Telescopio Carlos Sánchez operated on the island of Tenerife by the IAC in the Spanish Observatorio del Teide, and on observations made with the William Herschel Telescope (WHT proposal SW2019b06) and the Isaac Newton Telescope (INT) both operated on the island of La Palma by the Isaac Newton Group of Telescopes in the Spanish Observatorio del Roque de los Muchachos of the Instituto de Astrofísica de Canarias.
The work of M.P. and B.A.P was supported by a grant of the Romanian National Authority for Scientific Research and Innovation, CNCS -- UEFISCDI, project number PN-III-P1-1.1-TE-2019-1504.  The work of M.P. and O. V. was also partially supported by a grant of the Romanian National Authority for  Scientific  Research -- UEFISCDI, project number PN-III-P2-2.1-PED-2021-3625. 

%%%%%%%%%%%%%%%%%%%%%%%%%%%%%%%%%%%%%%%%%%%%%%%%%%
\section*{Data Availability}

%The inclusion of a Data Availability Statement is a requirement for articles published in MNRAS. Data Availability Statements provide a standardised format for readers to understand the availability of data underlying the research results described in the article. The statement may refer to original data generated in the course of the study or to third-party data analysed in the article. The statement should describe and provide means of access, where possible, by linking to the data or providing the required accession numbers for the relevant databases or DOIs.

Data available on request. The data underlying this article will be shared on reasonable request to the corresponding author.



\bibliographystyle{mnras}
\bibliography{2i_borisov} % if your bibtex file is called example.bib

%\begin{thebibliography}{99}
%\end{thebibliography}

%%%%%%%%%%%%%%%%%%%%%%%%%%%%%%%%%%%%%%%%%%%%%%%%%%

%%%%%%%%%%%%%%%%% APPENDICES %%%%%%%%%%%%%%%%%%%%%

\newpage
\appendix

\section{Observational logs}
This appendix contains the circumstances for TCS, INT and WHT observations. 

\onecolumn
\setlength\LTleft{0pt}
\setlength\LTright{0pt}
\begin{longtable}{@{\extracolsep{\fill}}c c c c c c c c c c c c c@{}}
	\caption{TCS observing campaign log of 2I/Borisov. The circumstances are presented in detail for each filter.}
	\label{TCS2I_log} \\
\hline\hline       
$N_{img}$ & Filter & UT & ZP (mag) & $\sigma$ (mag) & FWHM ($"$) & $r_h$ (au) & $\Delta$ (au) & T (mag) & $\alpha (^\circ)$ & $\theta_\odot\ (^\circ)$ & $\theta_{-v} (^\circ)$ & $\delta_\oplus (^\circ)$ \\
\hline   
98& g & 2019-10-03 5:41 & 27.665 & 0.027 & 2.4 & 2.48575 & 2.97410 & 17.8 & 18.5 & 294.2 & 329.2 & -11.9  \\
91 & r & 2019-10-03 5:40 & 27.678 & 0.023 & 2.1 & 2.48577 & 2.97413 & 17.8 & 18.5 & 294.2 & 329.2 & -11.9  \\
94 & z$_s$ & 2019-10-03 5:38 & 26.087 & 0.035 & 2.6 & 2.48578 & 2.97415 & 17.8 & 18.5 & 294.2 & 329.2 & -11.9  \\
54 & g & 2019-10-04 5:36 & 26.916 & 0.041 & 3.3 & 2.47294 & 2.95337 & 17.8 & 18.7 & 294.0 & 329.3 & -12.1  \\
72 & r & 2019-10-04 5:39 & 26.821 & 0.031 & 2.7 & 2.47292 & 2.95333 & 17.8 & 18.7 & 294.0 & 329.3 & -12.1  \\
58 & z$_s$ & 2019-10-04 5:37 & 25.333 & 0.017 & 3.3 & 2.47293 & 2.95335 & 17.8 & 18.7 & 294.0 & 329.3 & -12.1  \\
104 & g & 2019-10-05 5:43 & 27.809 & 0.055 & 2.3 & 2.46014 & 2.93255 & 17.8 & 18.9 & 293.8 & 329.4 & -12.3  \\
104 & r & 2019-10-05 5:42 & 27.698 & 0.036 & 2.7 & 2.46014 & 2.93256 & 17.8 & 18.9 & 293.8 & 329.4 & -12.3  \\
103 & z$_s$ & 2019-10-05 5:42 & 26.211 & 0.031 & 2.1 & 2.46015 & 2.93257 & 17.8 & 18.9 & 293.8 & 329.4 & -12.3  \\
83 & g & 2019-10-07 5:42 & 27.523 & 0.031 & 4.4 & 2.43505 & 2.89146 & 17.7 & 19.3 & 293.5 & 329.6 & -12.6  \\
83 & r & 2019-10-07 5:35 & 27.345 & 0.020 & 4.9 & 2.43510 & 2.89155 & 17.7 & 19.3 & 293.5 & 329.6 & -12.6  \\
81 & i & 2019-10-07 5:36 & 26.529 & 0.022 & 4.8 & 2.43509 & 2.89154 & 17.7 & 19.3 & 293.5 & 329.6 & -12.6  \\
83 & z$_s$ & 2019-10-07 5:39 & 26.756 & 0.027 & 4.0 & 2.43507 & 2.89150 & 17.7 & 19.3 & 293.5 & 329.6 & -12.6  \\
123 & g & 2019-10-12 5:29 & 27.794 & 0.046 & 2.9 & 2.37469 & 2.79086 & 17.6 & 20.3 & 292.7 & 330.1 & -13.6  \\
109 & r & 2019-10-12 5:25 & 27.781 & 0.031 & 1.9 & 2.37472 & 2.79092 & 17.6 & 20.3 & 292.7 & 330.1 & -13.6  \\
107 & i & 2019-10-12 5:24 & 26.922 & 0.046 & 1.9 & 2.37473 & 2.79093 & 17.6 & 20.3 & 292.7 & 330.1 & -13.6  \\
105 & z$_s$ & 2019-10-12 5:24 & 27.096 & 0.033 & 2.1 & 2.37473 & 2.79093 & 17.6 & 20.3 & 292.7 & 330.1 & -13.6  \\
97 & g & 2019-10-13 5:35 & 27.722 & 0.056 & 3.1 & 2.36296 & 2.77100 & 17.6 & 20.5 & 292.5 & 330.1 & -13.7  \\
100 & r & 2019-10-13 5:36 & 27.734 & 0.058 & 2.7 & 2.36295 & 2.77099 & 17.6 & 20.5 & 292.5 & 330.1 & -13.7  \\
100 & i & 2019-10-13 5:36 & 26.871 & 0.039 & 3.7 & 2.36296 & 2.77099 & 17.6 & 20.5 & 292.5 & 330.1 & -13.7  \\
101 & z$_s$ & 2019-10-13 5:36 & 27.010 & 0.036 & 3.1 & 2.36296 & 2.77099 & 17.6 & 20.5 & 292.5 & 330.1 & -13.7  \\
62 & g & 2019-10-14 5:53 & 27.793 & 0.127 & 2.6 & 2.33989 & 2.73160 & 17.5 & 20.9 & 292.2 & 330.3 & -14.1  \\
62 & r & 2019-10-14 5:51 & 27.767 & 0.053 & 2.6 & 2.33991 & 2.73163 & 17.5 & 20.9 & 292.2 & 330.3 & -14.1  \\
62 & i & 2019-10-14 5:53 & 26.891 & 0.056 & 2.3 & 2.33989 & 2.73160 & 17.5 & 20.9 & 292.2 & 330.3 & -14.1  \\
62 & z$_s$ & 2019-10-14 5:53 & 27.045 & 0.047 & 2.2 & 2.33989 & 2.73160 & 17.5 & 20.9 & 292.2 & 330.3 & -14.1  \\
104 & g & 2019-10-19 5:49 & 27.806 & 0.052 & 2.7 & 2.29587 & 2.65511 & 17.4 & 21.7 & 291.7 & 330.6 & -14.8  \\
85 & r & 2019-10-19 5:45 & 27.778 & 0.054 & 2.2 & 2.29590 & 2.65516 & 17.4 & 21.7 & 291.7 & 330.6 & -14.8  \\
88 & i & 2019-10-19 5:47 & 26.915 & 0.055 & 2.2 & 2.29589 & 2.65513 & 17.4 & 21.7 & 291.7 & 330.6 & -14.8  \\
93 & z$_s$ & 2019-10-19 5:46 & 27.018 & 0.038 & 2.8 & 2.29589 & 2.65514 & 17.4 & 21.7 & 291.7 & 330.6 & -14.8  \\
51 & g & 2019-10-20 5:39 & 27.565 & 0.106 & 2.3 & 2.28532 & 2.63648 & 17.4 & 21.9 & 291.6 & 330.6 & -14.9  \\
57 & r & 2019-10-20 5:41 & 27.494 & 0.058 & 3.3 & 2.28531 & 2.63646 & 17.4 & 21.9 & 291.6 & 330.6 & -14.9  \\
62 & i & 2019-10-20 5:42 & 26.608 & 0.060 & 3.1 & 2.28530 & 2.63645 & 17.4 & 21.9 & 291.6 & 330.6 & -14.9  \\
62 & z$_s$ & 2019-10-20 5:42 & 26.729 & 0.052 & 3.0 & 2.28530 & 2.63644 & 17.4 & 21.9 & 291.6 & 330.6 & -14.9  \\
81 & g & 2019-11-01 5:50 & 27.766 & 0.080 & 4.7 & 2.17094 & 2.42504 & 17.1 & 24.1 & 290.2 & 331.1 & -16.6  \\
84 & r & 2019-11-01 5:51 & 27.716 & 0.049 & 2.4 & 2.17093 & 2.42504 & 17.1 & 24.1 & 290.2 & 331.1 & -16.6  \\
84 & i & 2019-11-01 5:51 & 26.859 & 0.044 & 2.4 & 2.17094 & 2.42504 & 17.1 & 24.1 & 290.2 & 331.1 & -16.6  \\
83 & z$_s$ & 2019-11-01 5:51 & 26.977 & 0.047 & 4.4 & 2.17094 & 2.42504 & 17.1 & 24.1 & 290.2 & 331.1 & -16.6  \\
149 & g & 2019-11-02 5:31 & 27.699 & 0.037 & 2.2 & 2.16270 & 2.40894 & 17.1 & 24.3 & 290.1 & 331.2 & -16.7  \\
137 & r & 2019-11-02 5:28 & 27.654 & 0.018 & 2.0 & 2.16272 & 2.40898 & 17.1 & 24.3 & 290.1 & 331.2 & -16.7  \\
138 & i & 2019-11-02 5:28 & 26.776 & 0.017 & 2.1 & 2.16272 & 2.40898 & 17.1 & 24.3 & 290.1 & 331.2 & -16.7  \\
128 & z$_s$ & 2019-11-02 5:25 & 26.923 & 0.021 & 1.9 & 2.16273 & 2.40901 & 17.1 & 24.3 & 290.1 & 331.2 & -16.7  \\
106 & g & 2019-11-04 5:52 & 27.740 & 0.037 & 2.9 & 2.14646 & 2.37672 & 17.1 & 24.7 & 289.9 & 331.2 & -16.9  \\
98 & r & 2019-11-04 5:47 & 27.662 & 0.031 & 2.7 & 2.14649 & 2.37677 & 17.1 & 24.7 & 289.9 & 331.2 & -16.9  \\
98 & i & 2019-11-04 5:43 & 26.763 & 0.028 & 2.7 & 2.14651 & 2.37681 & 17.1 & 24.7 & 289.9 & 331.2 & -16.9  \\
92 & z$_s$ & 2019-11-04 5:41 & 26.945 & 0.033 & 2.5 & 2.14653 & 2.37684 & 17.1 & 24.7 & 289.9 & 331.2 & -16.9  \\
89 & g & 2019-11-06 5:51 & 27.621 & 0.045 & 3.1 & 2.13114 & 2.34566 & 17 & 25 & 289.8 & 331.2 & -17.1  \\
91 & r & 2019-11-06 5:47 & 27.524 & 0.027 & 3.2 & 2.13116 & 2.34570 & 17 & 25 & 289.8 & 331.2 & -17.1  \\
90 & i & 2019-11-06 5:41 & 26.737 & 0.026 & 2.8 & 2.13119 & 2.34576 & 17 & 25 & 289.8 & 331.2 & -17.1  \\
90 & z$_s$ & 2019-11-06 5:39 & 26.901 & 0.042 & 2.7 & 2.13120 & 2.34579 & 17 & 25 & 289.8 & 331.2 & -17.1  \\
89 & g & 2019-11-07 5:31 & 27.752 & 0.059 & 3.1 & 2.12388 & 2.33068 & 17 & 25.2 & 289.7 & 331.2 & -17.1  \\
83 & r & 2019-11-07 5:31 & 27.731 & 0.033 & 2.4 & 2.12388 & 2.33068 & 17 & 25.2 & 289.7 & 331.2 & -17.1  \\
83 & i & 2019-11-07 5:30 & 26.843 & 0.028 & 2.8 & 2.12389 & 2.33069 & 17 & 25.2 & 289.7 & 331.2 & -17.1  \\
83 & z$_s$ & 2019-11-07 5:29 & 27.025 & 0.028 & 2.3 & 2.12389 & 2.33070 & 17 & 25.2 & 289.7 & 331.2 & -17.1  \\
64 & g & 2019-11-10 5:35 & 27.646 & 0.029 & 2.9 & 2.10301 & 2.28654 & 16.9 & 25.6 & 289.5 & 331.1 & -17.3  \\
63 & r & 2019-11-10 5:34 & 27.603 & 0.026 & 2.6 & 2.10301 & 2.28654 & 16.9 & 25.6 & 289.5 & 331.1 & -17.3  \\
61 & i & 2019-11-10 5:29 & 26.765 & 0.032 & 2.5 & 2.10303 & 2.28659 & 16.9 & 25.6 & 289.5 & 331.1 & -17.3  \\
58 & z$_s$ & 2019-11-10 5:27 & 26.934 & 0.037 & 2.4 & 2.10304 & 2.28661 & 16.9 & 25.6 & 289.5 & 331.1 & -17.3  \\
87 & g & 2019-11-11 5:31 & 27.777 & 0.064 & 2.4 & 2.09649 & 2.27237 & 16.9 & 25.8 & 289.4 & 331.1 & -17.4  \\
83 & r & 2019-11-11 5:30 & 27.703 & 0.033 & 2.2 & 2.09649 & 2.27238 & 16.9 & 25.8 & 289.4 & 331.1 & -17.4  \\
85 & i & 2019-11-11 5:31 & 26.848 & 0.057 & 2.1 & 2.09649 & 2.27238 & 16.9 & 25.8 & 289.4 & 331.1 & -17.4  \\
83 & z$_s$ & 2019-11-11 5:30 & 26.999 & 0.027 & 2.0 & 2.09649 & 2.27238 & 16.9 & 25.8 & 289.4 & 331.1 & -17.4  \\
56 & r & 2019-11-17 5:40 & 27.644 & 0.059 & 2.3 & 2.06182 & 2.19258 & 16.8 & 26.7 & 289.2 & 330.8 & -17.6  \\
56 & i & 2019-11-17 5:36 & 26.786 & 0.036 & 2.3 & 2.06183 & 2.19262 & 16.8 & 26.7 & 289.2 & 330.8 & -17.6  \\
52 & z$_s$ & 2019-11-17 5:34 & 26.940 & 0.040 & 2.1 & 2.06184 & 2.19264 & 16.8 & 26.7 & 289.2 & 330.8 & -17.6  \\
38 & g & 2019-11-18 6:05 & 27.770 & 0.031 & 2.6 & 2.05673 & 2.18003 & 16.8 & 26.8 & 289.2 & 330.8 & -17.6  \\
57 & r & 2019-11-18 6:10 & 27.705 & 0.060 & 1.8 & 2.05672 & 2.17999 & 16.8 & 26.8 & 289.2 & 330.8 & -17.6  \\
56 & i & 2019-11-18 6:10 & 26.821 & 0.043 & 2.5 & 2.05672 & 2.17999 & 16.8 & 26.8 & 289.2 & 330.8 & -17.6  \\
58 & z$_s$ & 2019-11-18 6:09 & 26.997 & 0.050 & 1.8 & 2.05672 & 2.18001 & 16.8 & 26.8 & 289.2 & 330.8 & -17.6  \\
31 & g & 2019-11-22 6:20 & 27.718 & 0.039 & 2.9 & 2.03903 & 2.13344 & 16.7 & 27.3 & 289.1 & 330.4 & -17.5  \\
31 & r & 2019-11-22 6:20 & 27.692 & 0.031 & 2.5 & 2.03903 & 2.13344 & 16.7 & 27.3 & 289.1 & 330.4 & -17.5  \\
31 & i & 2019-11-22 6:20 & 26.806 & 0.037 & 3.1 & 2.03902 & 2.13343 & 16.7 & 27.3 & 289.1 & 330.4 & -17.5  \\
31 & z$_s$ & 2019-11-22 6:20 & 26.992 & 0.027 & 2.7 & 2.03903 & 2.13344 & 16.7 & 27.3 & 289.1 & 330.4 & -17.5  \\
97 & g & 2019-11-25 6:02 & 27.802 & 0.046 & 2.5 & 2.02827 & 2.10177 & 16.7 & 27.6 & 289.1 & 330.1 & -17.4  \\
97 & r & 2019-11-25 5:57 & 27.736 & 0.037 & 1.9 & 2.02828 & 2.10180 & 16.7 & 27.6 & 289.1 & 330.1 & -17.4  \\
96 & i & 2019-11-25 5:54 & 26.867 & 0.039 & 2.5 & 2.02829 & 2.10182 & 16.7 & 27.6 & 289.1 & 330.1 & -17.4  \\
96 & z$_s$ & 2019-11-25 5:54 & 27.020 & 0.033 & 1.9 & 2.02829 & 2.10182 & 16.7 & 27.6 & 289.1 & 330.1 & -17.4  \\
145 & g & 2019-12-02 5:59 & 26.952 & 0.050 & 3.6 & 2.01145 & 2.03821 & 16.6 & 28.2 & 289.3 & 329.2 & -16.7  \\
144 & r & 2019-12-02 5:56 & 26.782 & 0.026 & 4.9 & 2.01145 & 2.03822 & 16.6 & 28.2 & 289.3 & 329.2 & -16.7  \\
143 & i & 2019-12-02 5:54 & 26.037 & 0.031 & 3.5 & 2.01146 & 2.03824 & 16.6 & 28.2 & 289.3 & 329.2 & -16.7  \\
144 & z$_s$ & 2019-12-02 5:51 & 26.119 & 0.042 & 4.3 & 2.01146 & 2.03825 & 16.6 & 28.2 & 289.3 & 329.2 & -16.7  \\
294 & g & 2019-12-04 5:34 & 27.271 & 0.031 & 2.3 & 2.00885 & 2.02299 & 16.6 & 28.3 & 289.4 & 328.8 & -16.5  \\
294 & r & 2019-12-04 5:32 & 27.233 & 0.021 & 2.9 & 2.00885 & 2.02300 & 16.6 & 28.3 & 289.4 & 328.8 & -16.5  \\
294 & i & 2019-12-04 5:27 & 26.375 & 0.024 & 2.5 & 2.00886 & 2.02302 & 16.6 & 28.3 & 289.4 & 328.8 & -16.5  \\
219 & z$_s$ & 2019-12-04 5:27 & 26.575 & 0.027 & 5.2 & 2.00886 & 2.02302 & 16.6 & 28.3 & 289.4 & 328.8 & -16.5  \\
52 & g & 2019-12-08 6:05 & 27.614 & 0.051 & 3.0 & 2.00655 & 1.99595 & 16.6 & 28.5 & 289.7 & 328.1 & -15.8  \\
50 & r & 2019-12-08 5:59 & 27.581 & 0.038 & 2.5 & 2.00655 & 1.99598 & 16.6 & 28.5 & 289.7 & 328.1 & -15.8  \\
52 & i & 2019-12-08 6:00 & 26.738 & 0.037 & 2.5 & 2.00655 & 1.99597 & 16.6 & 28.5 & 289.7 & 328.1 & -15.8  \\
49 & z$_s$ & 2019-12-08 5:59 & 26.909 & 0.039 & 2.3 & 2.00655 & 1.99598 & 16.6 & 28.5 & 289.7 & 328.1 & -15.8  \\
27 & g & 2019-12-10 6:23 & 27.711 & 0.048 & 2.4 & 2.00689 & 1.98436 & 16.5 & 28.6 & 289.9 & 327.6 & -15.5  \\
26 & r & 2019-12-10 6:24 & 27.669 & 0.039 & 2.2 & 2.00689 & 1.98436 & 16.5 & 28.6 & 289.9 & 327.6 & -15.5  \\
29 & i & 2019-12-10 6:22 & 26.825 & 0.038 & 2.2 & 2.00689 & 1.98436 & 16.5 & 28.6 & 289.9 & 327.6 & -15.5  \\
29 & z$_s$ & 2019-12-10 6:23 & 26.985 & 0.043 & 1.9 & 2.00689 & 1.98436 & 16.5 & 28.6 & 289.9 & 327.6 & -15.5  \\
32 & g & 2019-12-13 5:42 & 27.686 & 0.052 & 2.4 & 2.00923 & 1.96962 & 16.5 & 28.6 & 290.3 & 326.9 & -14.9  \\
25 & r & 2019-12-13 5:40 & 27.624 & 0.022 & 2.3 & 2.00923 & 1.96962 & 16.5 & 28.6 & 290.3 & 326.9 & -14.9  \\
28 & i & 2019-12-13 5:41 & 26.815 & 0.043 & 2.5 & 2.00923 & 1.96962 & 16.5 & 28.6 & 290.3 & 326.9 & -14.9  \\
25 & z$_s$ & 2019-12-13 5:40 & 26.982 & 0.040 & 2.2 & 2.00923 & 1.96962 & 16.5 & 28.6 & 290.3 & 326.9 & -14.9  \\
\hline                  
\end{longtable}

%%%%%%%%%%%%%%%%%%%%%%%%%%%%%%%%%%%%%%%%%%%%%%%%%%

\setlength\LTleft{0pt}
\setlength\LTright{0pt}
\begin{longtable}{@{\extracolsep{\fill}}c c c c c c c c c c c c c@{}}
	\caption{INT observing campaign log of 2I/Borisov. The circumstances are presented in detail for each filter.}
	\label{INT2I_log} \\
\hline\hline       
$N_{img}$ & Filter & UT & ZP (mag) & $\sigma (mag)$ & FWHM ($"$) & $r_h$ (au) & $\Delta$ (au) & T (mag) & $\alpha (^\circ)$ & $\theta_\odot\ (^\circ)$ & $\theta_{-v} (^\circ)$ & $\delta_\oplus (^\circ)$ \\
\hline   
3& B & 2019-10-04 5:11 & 28.135 & 0.065 & 2.3 & 2.47315 & 2.95372 & 17.8 & 18.7 & 294.0 & 329.3 & -12.1  \\
3 & i & 2019-10-04 5:17 & 28.237 & 0.048 & 2.0 & 2.47310 & 2.95364 & 17.8 & 18.7 & 294.0 & 329.3 & -12.1  \\
3 & r & 2019-10-04 5:14 & 28.531 & 0.048 & 2.0 & 2.47313 & 2.95368 & 17.8 & 18.7 & 294.0 & 329.3 & -12.1  \\
12 & V & 2019-10-04 5:14 & 28.419 & 0.038 & 2.2 & 2.47313 & 2.95368 & 17.8 & 18.7 & 294.0 & 329.3 & -12.1  \\
4 & B & 2019-10-05 4:53 & 28.286 & 0.095 & 2.2 & 2.46058 & 2.93328 & 17.8 & 18.9 & 293.8 & 329.4 & -12.3  \\
4 & i & 2019-10-05 4:58 & 28.325 & 0.058 & 1.8 & 2.46052 & 2.93319 & 17.8 & 18.9 & 293.8 & 329.4 & -12.3  \\
4 & r & 2019-10-05 4:55 & 28.614 & 0.058 & 1.8 & 2.46055 &2.93323 & 17.8 & 18.9 & 293.8 & 329.4 & -12.3  \\
16 & V & 2019-10-05 4:57 & 28.487 & 0.048 & 2.0 & 2.46053 & 2.93321 & 17.8 & 18.9 & 293.8 & 329.4 & -12.3  \\
3 & B & 2019-10-06 5:17 & 28.374 & 0.069 & 2.2 & 2.44775 & 2.91232 & 17.8 & 19.1 & 293.7 & 329.5 & -12.5  \\
3 & i & 2019-10-06 5:22 & 28.412 & 0.060 & 1.9 & 2.44770 & 2.91223 & 17.8 & 19.1 & 293.7 & 329.5 & -12.5  \\
3 & r & 2019-10-06 5:19 & 28.688 & 0.044 & 2.0 & 2.44772 & 2.91227 & 17.8 & 19.1 & 293.7 & 329.5 & -12.5  \\
12 & V & 2019-10-06 5:20 & 28.581 & 0.042 & 2.1 & 2.44772 & 2.91227 & 17.8 & 19.1 & 293.7 & 329.5 & -12.5  \\
4 & B & 2019-10-18 5:07 & 28.131 & 0.076 & 1.9 & 2.37469 & 2.79086 & 17.5 & 21.5 & 291.8 & 330.5 & -14.6  \\
4 & i & 2019-10-18 5:13 & 28.005 & 0.080 & 1.8 & 2.37472 & 2.79092 & 17.5 & 21.5 & 291.8 & 330.5 & -14.6  \\
4 & r & 2019-10-18 5:10 & 28.386 & 0.012 & 1.9 & 2.37473 & 2.79093 & 17.5 & 21.5 & 291.8 & 330.5 & -14.6  \\
16 & V & 2019-10-18 5:13 & 28.303 & 0.094 & 1.8 & 2.37473 & 2.79093 & 17.5 & 21.5 & 291.8 & 330.5 & -14.6  \\
2 & B & 2019-10-30 5:49 & 28.808 & 0.049 & 2.6 & 2.18821 & 2.45834 & 17.2 & 23.8 & 290.4 & 331.1 & -16.3  \\
2 & i & 2019-10-30 5:57 & 28.737 & 0.059 & 1.7 & 2.18816 & 2.45825 & 17.2 & 23.8 & 290.4 & 331.1 & -16.3  \\
2 & r & 2019-10-30 5:53 &29.074 & 0.067 & 2.4 & 2.18818 & 2.45829 & 17.2 & 23.8 & 290.4& 331.1 & -16.3  \\
8 & V & 2019-10-30 5:52 & 28.957 & 0.055 & 2.2 & 2.18819 & 2.45831 & 17.2 & 23.8 & 290.4 & 331.1 & -16.3  \\

\hline                  
\end{longtable}
\textbf{ } \\
$N_{img}$ - the number of images used in stacking, as we removed a part of TCS images because of tracking errors or unfavourable circumstances (i.e. star occultations) \\
UT - Universal Time (yyyy-MM-dd HH:mm) \\
ZP - the zeropoint (calibration via Pan-STARRS catalog), in magnitudes\\
$\sigma$ - the zeropoint uncertainty, in magnitudes (by propagation becomes the uncertainty of the estimated magnitude)\\
FWHM - The Full Width at Half-Maximum of the Point Spread Function , in arcseconds\\
$r_h$ - heliocentric distance, in astronomical units \\
$\Delta$ - distance to the Earth, in astronomical units \\
$T$ - estimated total magnitude of the comet, in magnitudes\\
$\alpha$ - phase angle, in degrees \\
$\theta_\odot$ - position angle of the anti-solar vector, in degrees \\
$\theta_{-v}$ - position angle of the anti-velocity vector, in degrees \\
$\delta_\oplus$ - orbital plane angle measuring the comet elevation above the ecliptic as seen by a terrestrial observer, in degrees
%%%%%%%%%%%%%%%%%%%%%%%%%%%%%%%%%%%%%%%%%%%%%%%%%%
\setlength\LTleft{0pt}
\setlength\LTright{0pt}
\begin{longtable}{@{\extracolsep{\fill}}c c c c c c c c c @{}}
	\caption{WHT observing circumstances of 2I/Borisov. }
	\label{WHT2I_log} \\
\hline\hline       
$N_{img}$ & Instrument & UT & Exposure (s)  & $r_h$ (au) & $\Delta$ (au) & T (mag) & $\alpha (^\circ)$  \\
\hline   
13& ISIS red arm & 2019-11-27 5:40 & 4 x 1200 &  2.02519 & 2.09191 & 16.6 & 27.6  \\
13& ISIS blue arm & 2019-11-27 5:40 & 4 x 1200 &  2.02519 & 2.09191 & 16.6 & 27.6  \\

\hline                  
\end{longtable}

\twocolumn
\section{Dust Model}
\label{apx:dust}
The model described in this section for determining the dust production rate has been proposed in \cite{2012Icar..221..721F}. The continuum spectra detected by observers on Earth from a comet originates from sunlight that reflects off the cometary dust. When a comet becomes active, gas released by the nucleus carries dust particles until they reach their terminal velocity. According to direct simulations using the Monte Carlo approach, the gas-dust decoupling happens around 10-20 km from a 2 km nucleus (\cite{2011ApJ...732..104T}), which is relatively short compared to the coma's size. Therefore, the model is based on the following assumptions:
\begin{itemize}
    \item the outflow velocity is constant for dust particles of same size, $a$, denoted as $v_d = v_d(a)$. The above mentioned Monte Carlo simulations show that the terminal speed is about $v_d(a) \approx 0.1423 \dfrac{m}{s} \Big(\dfrac{a}{\SI{1}{\meter}}\Big)^{-0.5}$
    \item the distribution and outflow of cometary dust are spherically symmetric, and there is no dust production or destruction inside the coma.
\end{itemize} 

The coma length scale in the sunward direction, as restricted by the solar radiation pressure \citep{1991ASSL..167...19J}, can be estimated as 
\begin{equation}
  X_R=\frac{v_d^2r_h^2}{2\beta GM_\odot}  
  \label{eq:xr}
\end{equation}
where $r_h$ is the heliocentric distance, and $\beta$ is the radiation pressure factor, which is roughly equal to $\frac{1 \ \mu m}{a}$ for dielectric spheres \citep{1983asls.book.....B}. Thus, Eq. \ref{eq:xr} leads to $v_d(a)=k_d a^{-0.5}$, and a linear relationship between the coma length scale and the square of the heliocentric distance.

The dust production rate ($Q_d$) represents the total number of particles generated by the nucleus per unit time. During a time interval $\Delta t$, $\Delta N = Q_d \Delta t$ particles are created, which will travel through the volume $\Delta V = 4\pi r^ 2 \Delta r$ at a radial distance $r$ from the nucleus with a constant outflow velocity of $v_d$. Therefore, $\Delta r = v_d \Delta t$ and the dust volume density of particles as a function of $r$ is
\begin{equation}
 n_{vol}=\frac{\Delta N}{\Delta V}=\frac{Q_d}{4\pi v_d r^2}   
 \label{nvol}
\end{equation}

The surface density as a density of a column of dust along the line of sight at a distance $\rho$ from the nucleus and projected in the sky plane can be calculated by integrating Eq. \ref{nvol} along the line of sight, $z$. Substituting $r^2=\rho^2 + z^2$, and assuming that $\rho << z$ one gets
\begin{equation}
n_{col}(\rho)=\int_{-\infty}^{\infty} \frac{Q_d}{4\pi v_d} \frac{1}{\rho^2 + z^2}  dz = \frac{Q_d}{4v_d\rho}
\end{equation}
Then, the number of dust particles is derived by integrating over the considered aperture radius $R$, such that
\begin{equation}
    N = \int_0^R n_{col}(\rho) 2\pi \rho d\rho = \frac{Q_d\pi R}{2v_d}
    \label{eq:N}
\end{equation}

The terminal velocity of dust outflow and the scattering efficiency depend on the size of the dust particles. Thus, in order to determine macroscopic quantities such as mass or dust production rate one needs to make an assumption about the size distribution of the particles. For this, the size distribution formula proposed in \cite{2012Icar..221..721F} is
\begin{equation}
    \frac{dQ}{da} = g_0 e^{-\frac{a_0}{a}}\Big(\frac{a_0}{a}\Big)^\alpha
\label{eq:dust_dist}
\end{equation}

where $g_0$ is a normalisation constant and, $a_0$ and $\alpha$ are free constants. The most probable size given by the peak of the size distribution is $a_p=\frac{a_0}{\alpha}$.

The equivalent grain density distribution derives from the equations \ref{eq:N} and \ref{eq:dust_dist} as
\begin{equation}
    \frac{dN}{da} = \frac{\pi R}{2v_d} g_0 e^{-\frac{a_0}{a}}\Big(\frac{a_0}{a}\Big)^\alpha
\end{equation}
Hence, the total number of dust grains, mass, and dust production rate will be
\begin{equation}
    N=\frac{\pi R}{2} \int_{a_{min}}^{a_{max}} g_0 \frac{1}{v_d} e^{-\frac{a_0}{a}}\Big(\frac{a_0}{a}\Big)^\alpha da
\end{equation}
\begin{equation}
    M= \frac{\pi R}{2} \int_{a_{min}}^{a_{max}} g_0 \rho(a) \frac{4 \pi a^3}{3v_d} e^{-\frac{a_0}{a}}\Big(\frac{a_0}{a}\Big)^\alpha da
\end{equation}
\begin{equation}
    Q_d=\int_{a_{min}}^{a_{max}} g_0 e^{-\frac{a_0}{a}}\Big(\frac{a_0}{a}\Big)^\alpha da
\end{equation}
Further, one needs to know the scattering efficiency in order to determine the total scattering cross-section of the comet. The scattering efficiency $q_s(a, \lambda)$ is computed by Mie scattering theory \cite{1908AnP...330..377M} \cite{1974SSRv...16..527H} and it depends on the wavelength and the optical properties of the dust. To compute these scattering coefficients we use PyMieScatt, a Python software developed by \cite{2018JQSRT.205..127S}, which applies the Mie solution of the Maxwell's equation to find the scattering coefficient $q_s(a, \lambda)$ for a spherical particle of size $a$ at wavelength $\lambda$. We estimate these coefficients for a specific refractive index $m=n+ik$ considering the effective wavelength of each band. 

The total scattering cross-section, $C_e$, is found by integrating over the scattering area of each dust particle, $\sigma = q_s(a, \lambda)\pi a^2$. Namely,
\begin{equation}
    C_e=\frac{\pi R}{2} \int_{a_{min}}^{a_{max}} g_0 \frac{q_s(a, \lambda)}{v_d} \pi a^2 e^{-\frac{a_0}{a}}\Big(\frac{a_0}{a}\Big)^\alpha da
    \label{eq:Ce_dust_dist}
\end{equation}

The maximum size of the grains depends on the gravitational force exerted by the nucleus and the gas drag force entertaining the dust. \cite{2011ApJ...732..104T} One can write the second law of mechanics for a grain having the mass $m=\rho \frac{4\pi a^3}{3}$ as
\begin{equation}
m\dot{\vec{v}}_d = \frac{1}{2} C_d \pi a^2  \rho_g (\vec{v}_{g} - \vec{v}_d) \mid \vec{v}_{g} - \vec{v}_d \mid - \frac{GM_nm}{r^2} \frac{\vec{r}}{r}   
\label{eq:second_law}
\end{equation}
where $\vec{v_g}$ is the gas velocity, $G$ is the gravitational constant, $M_n=\rho_n \frac{4 \pi r_n^3}{3}$ is the nucleus mass, $\rho_g$ is the gas density and $C_d$ is the drag coefficient. One can set $a_d = 0$ and $r=r_n$ (nucleus radius) to find the size limit allowed by the gravitational force exerted by the nucleus. Thus,
\begin{equation}
a_{max} = \frac{3C_d r_n^2  v_g^2 \rho_{g}}{8 GM_n \rho} 
\label{eq:max_size}
\end{equation}
A accepted value for the drag coefficient of cometary gas is $C_d=2$ \cite{1986RvGeo..24..667G}. The outflow gas velocity can be determined from the gas temperature,
\begin{equation}
    \frac{1}{2} \mu v_g^2 = \frac{3}{2} k_B T
\end{equation}
where $\mu$ is the molecular mass and $k_B$ the Boltzmann constant. Hence, equation \ref{eq:max_size} can be rewritten as
\begin{equation}
a_{max} = \frac{27C_d k_B T \rho_{g}}{32\pi Gr_n \rho_n \rho\mu} 
\label{eq:max_size2}
\end{equation}

\section{Haser model}
\label{apx:haser}
To compute the production rate (Q) of molecules in the comet's coma, we measure the flux $F_i=L_i/4\pi\Delta^2$ of a specific molecular emission band $i$. Then, the number of molecules within the observing aperture is computed using the fluorescence efficiency, $L/N$, called also "g-factor". 

Using Haser model \citep{Haser_1957, Cochran_1985}, we extrapolate the total number of molecules in the entire coma, by considering the following density distribution of the molecules,
\begin{equation}
    n=\frac{Q}{4 \pi r^2 v}\frac{\lambda_D\lambda_P}{\lambda_P-\lambda_D} \Large(e^{-r/\lambda_P}-e^{-r/\lambda_D}\Large)
\end{equation}
One can notice that this is a radial distribution, depending on a radial position, $r$, outflow velocity $v$, and the parent and daughter scale lengths, $\lambda_{P,D}$.
The production rate is then derived ensuring static equilibrium between destruction and production rates,
\begin{equation}
    Q_i=\frac{N_i}{\tau_i}=\frac{L_i}{g_i\tau_i}=\frac{4\pi\Delta^2F_i}{g_i\tau_i}
\end{equation}
where $\tau_i$ is the lifetime of the daughter molecules, and it is equal to the ratio between the daughter scale length, $\lambda_D$, and the outflow velocity, which can be estimated as $v=0.58/\sqrt{r_h}$ km/s \citep{1982come.coll...85D}.
The values for the g-factors and scale lengths are provided in Table~\ref{tbl:haser} for the standard heliocentric distance of 1 au and the heliocentric distance at the time we observed the comet at 2 au.  The scale lengths at 1 au for CN are retrieved from \cite{1995Icar..118..223A}, whereas for NH$_2$ from \cite{Fink_1996}. Regarding the fluorescence efficiencies, we use those reported by \cite{2010AJ....140..973S} for CN and by \cite{kawakita_2001} for NH$_2$.  
\begin{table}
\centering
\begin{tabular}{ccccc}
\hline
Species                                                                     & r (au) & $g$ (erg/s)           & $\lambda_P$ (km) & $\lambda_D$ (km) \\ \hline
\multirow{2}{*}{\begin{tabular}[c]{@{}c@{}}CN\\ (0-0), 388 nm\end{tabular}} & 1.0    & $3.6\times 10^{-13}$               & 13000            & 210000           \\
                                                                            & 2.0    & $8.7\times 10^{-14}$               & 53500            & 865000           \\
\multirow{2}{*}{\begin{tabular}[c]{@{}c@{}}NH$_2$\\ (0, 8, 0)\end{tabular}} & 1.0    & $4.44 \times 10^{-15}$ & 4900             & 62000 \\
                                                                            & 2.0    & $1.77\times 10^{-15}$              & 20000            & 255000           \\ \hline
\end{tabular}
\caption{Haser model properties of CN and NH$_2$ species detected in this work}
\label{tbl:haser}
\end{table}

\section{Composite images with 2I/Borisov}
In this appendix we provide the composite images in r-band for the night observed with TCS.

\begin{figure*}
\includegraphics[width=0.85\linewidth]{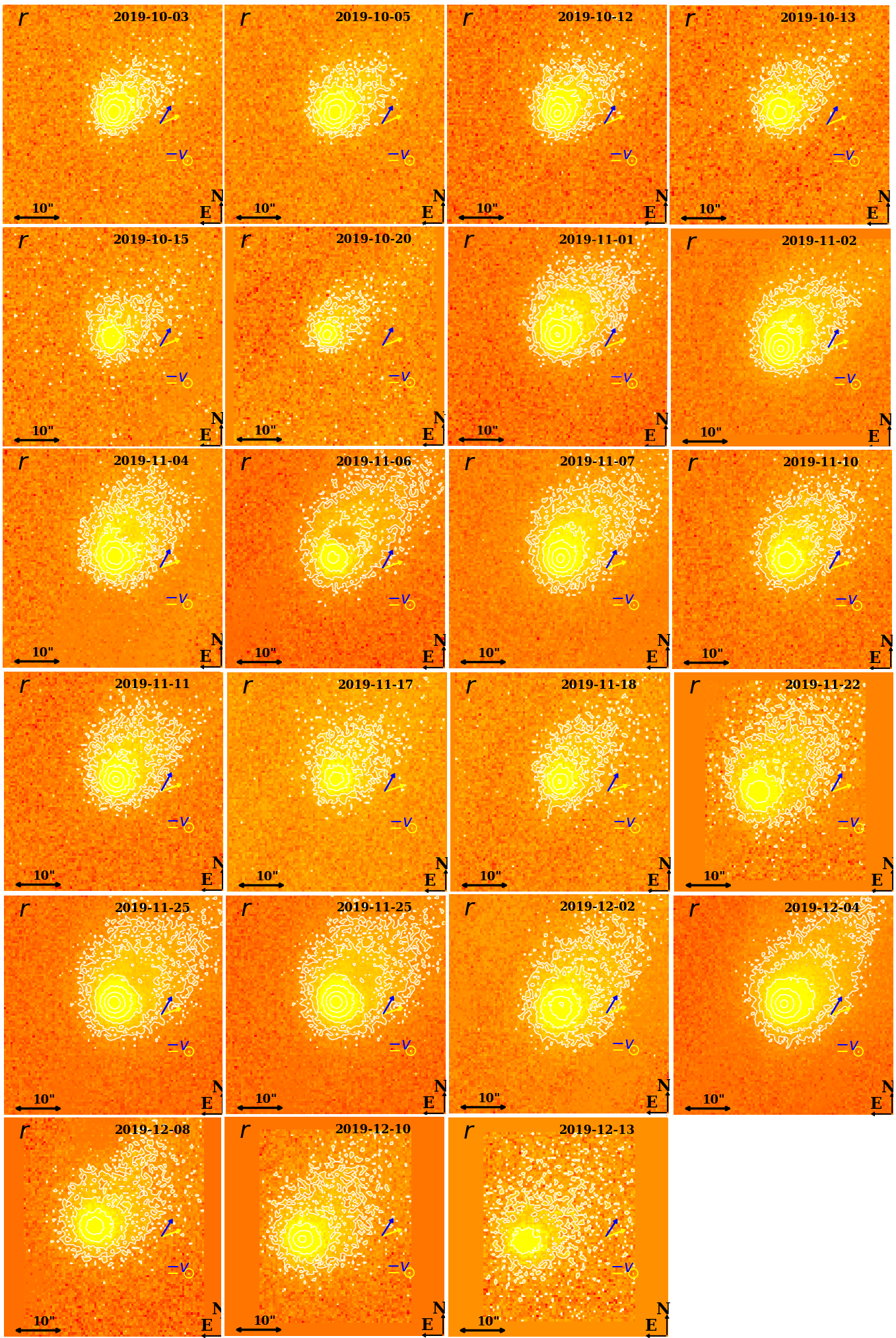}
\caption{Composite images of 2I in r-band with the comet during the observational campaign. They are obtained using the TCS data by tracking on the comet's motion and stacking. The number of co-added frames can be checked in the observing log in Table~\ref{TCS2I_log}. North (N) and east (E) directions of the equatorial coordinate system, and the image scale of 10 arcseconds are illustrated. The anti-solar (\textbf{-$\mathbf{\odot}$}) and anti-velocity (\textbf{-v}) orientations retrieved with JPL Horizons are drawn with yellow, and blue, respectively.}
\end{figure*}
% Don't change these lines
\bsp	% typesetting comment
\label{lastpage}
\end{document}